\documentstyle[aps,preprint]{revtex}
\tighten
\begin{document}
\draft
\title
{From microscopic description to statistical mechanics of Cu-O chain fragments
in the high-T$_{\rm c}$
superconductors REBa$_2$Cu$_3$O$_{6+{\rm x}}$}
\author{
H\aa rek Haugerud}
\address{University of Oslo, Department of
Physics, P.O. Box 1048 - Blindern 0316 Oslo, Norway}
\author{Gennadi Uimin\cite{gu}}
\address{Institut f\"ur Theoretische Physik, Universit\"at zu K\"oln,
D-50937 K\"oln, Germany\\
and\\
D\'epartement de Recherche Fondamentale sur la Mati\`ere Condens\'ee,
SPSMS/MDN, CENG, 17, rue des Martyrs, F-38054 Grenoble Cedex 9, France}
\author{Walter Selke}
\address{Institut f\"ur Theoretische Physik B, Technische Hochschule,
D-52056 Aachen, Germany}
\maketitle

\begin{abstract}
The oxygen deficient CuO$_{\rm x}$ planes of the high-T$_{\rm c}$
superconductors
REBa$_2$Cu$_3$O$_{6+{\rm x}}$ (RE denotes a rare-earth atom or yttrium)
are studied. Starting from the Emery model for the Cu-O chain fragments,
an {\it extended} ASYNNNI model is proposed to describe the statistical
mechanics of the CuO$_{\rm x}$ planes. The model is analysed by using
Monte Carlo techniques, computing especially the charge transfer
to the CuO$_2$ planes regulating the high-T$_{\rm c}$
superconductivity, and the
concentration of differently coordinated Cu ions.
Results are compared to experimental data for
various RE cases.
\end{abstract}
\vspace{4cm}
\pacs{Keywords: REBCO compounds, Cu-O chains, Emery model, ASYNNNI model}
\section{Introduction}
The high-T$_{\rm c}$ superconductors
REBa$_2$Cu$_3$O$_{6+{\rm x}}$ (REBCO), where RE denotes yttrium or one of the
rare-earth atoms Yb, Er, Y, Sm, Nd, or La, have been studied extensively,
both theoretically and experimentally. Among others, the
impact of the oxygen deficient  CuO$_{\rm x}$ (sometimes
called Cu(1)) planes on superconductivity has
been investigated. In particular, the charge transfer from the Cu(1)
planes to the superconducting CuO$_2$ (or Cu(2)) planes has been
shown to play a crucial role.

The Cu(1) planes consist of linear Cu--O--$\,\cdots\,$--O--Cu
chain fragments (CFs). Each Cu atom is strongly coupled to the two
apical oxygen atoms in adjacent Ba-O planes. Depending on the number, $l$,
of neighboring oxygen atoms in a CF, Cu atoms may be
$(l\!+\!2)$-fold coordinated. Obviously, a threefold coordinated
Cu atom, Cu$^{(3)}$, is located at the end of a chain fragment; fourfold
coordination (Cu$^{(4)}$) is realized for a copper atom in the CF, while
an isolated Cu atom is twofold coordinated, Cu$^{(2)}$.
The concentration of differently coordinated Cu atoms has been determined
experimentally by means of NMR-NQR \cite{luetg1,luetg2} and x-ray
absorption spectroscopy (XAS) \cite{tranq,tolen}.
Using the soft XAS method, it is possible to estimate the concentration of
oxygen holes in sufficiently long Cu-O chains. For example, for untwinned
YBa$_2$Cu$_3$O$_7$, it has been found to be about 30\% \cite{krol}.

$T_c$, the superconducting phase transition temperature, as a function
of oxygen content $x$ is known to display the
prominent 60 K and 90 K plateaus for RE = Yb, Y and Er \cite{BHHKW90}, i.e.,
the compounds of comparatively small RE ionic radii. The analog of the
60 K plateau is less pronounced for RE = Gd \cite{veal_4}, and the plateaus 
are practically absent for the REBCO compounds with
RE = Nd \cite{BHHKW90,veal_4} and La \cite{lindemer}.
The distribution of chain lengths in the CF ensemble allows one to
estimate the concentration of holes leaving the chains, $n_h$,
which is usually assumed to be
proportional to $T_c$, unless one is in the underdoping
($n_h\!<\!0.05$) and overdoping
($n_h\!>\!0.25$) regimes. Underdoping corresponds to localization of
transferred holes in CuO$_2$ planes, while overdoping holds
near the maximum of $T_c(n_h)$ (see \cite{lant_1} and  \cite{lant_2}).

$T_c$ depends not only on $x$ but also on the temperature, $T_q$, at
which the Cu(1) planes have been effectively equilibrated experimentally.
$T_q$ is typically the annealing temperature for rapid quenches, or
the room temperature, $T_R$, if the cooling was done slowly; below
$T_R$, the mobility of the oxygen atoms in the CuO$_x$ planes
is greatly reduced.
Varying $T_q$ up to 350$^{\circ}$C in experiments on YBCO,
Veal and co-workers \cite{veal_1} showed that the 60 K plateau is sensitive
to changes of $T_q$. In fact, the plateau becomes less pronounced with
increasing $T_q$.
A subsequent experiment \cite{gant} confirmed these results, and demonstrated
that the various annealing temperatures $T_q$ are accompanied by
rearrangements in the ensemble of chain fragments.
Note, that the temperature considered in our theoretical analysis of the
Cu--O chains corresponds to $T_q$ in experiments.

The Cu-O chains in the Cu(1) planes may form differently ordered
phases, the orthorhombic, Ortho-I, Ortho-II, and probably
Ortho-III (see, for instance, \cite{zeis,plakh_1,plakh_2,hohl_2,and_1}, as
well as the
tetragonal phases, which have been identified
experimentally. The perfect Ortho-I and Ortho-II structures together with
a typical arrangement of oxygen atoms in the
tetragonal structure are shown in Figure \ref{figure1}.

The transformation from the semiconducting to the metal state
in YBCO is usually ascribed
to the tetragonal--orthorhombic transition. The underlying
microscopic mechanism seems to be the
delocalization of holes in the  CuO$_2$ planes resulting
in a change of the oxygen hole chemical potential, which, in turn,
influences the structural properties \cite{gaw_2}.

Some years ago, a lattice--gas model, the asymmetric
next-nearest-neighbor Ising (ASYNNNI) model, has been
proposed for describing the oxygen ordering in 
the Cu(1) planes of YBCO \cite{font_1}.
It is based on three effective coupling constants as shown in Figure
\ref{figure2}. Two of these couplings,
the nearest neighbor oxygen repulsion $V_1$ ($>0$) and the
next-nearest neighbor attraction $V_2$ ($<0$) mediated by a copper atom,
favor Ortho-I-like configurations. The third coupling, $V_3$, the direct
interaction between next-nearest oxygen atoms without intermediate Cu atom,
is supposed to be repulsive, thereby favoring formation of
Ortho-II-like configurations. The phase diagram of the model is sketched
in Figure \ref{figure3}. As noted before, when comparing to experiments,
the temperature should be interpreted as $T_q$.

The phase diagram of the ASYNNNI model has been analyzed using various
numerical approaches,
such as cluster variation methods \cite{wille,font_2,OLDSET2},
Monte Carlo simulations for static \cite{and_1,selk,fiig1,fiig2} as well
as dynamic properties \cite{kinet}, and
transfer-matrix finite-size techniques \cite{eins}. Among the approximate
analytical approaches, one may mention a low-temperature expansion of the free
energy \cite{oit} and high-temperature expansions \cite{uim_int,uim_pr}.
The model allows to reproduce remarkably well the main features of the
experimental phase diagram for YBCO.
As a note of caution, attention may be drawn, however, to
experimental findings providing evidence for a possible discrepancy in the
location of the phase boundary of the Ortho-II phase
\cite{gant,veal_3,OIITc,schw}. According to these experiments, the
disordering of the Ortho-II phase, 
oxygen content $x$ close to 0.5, occurs at about 450 K.  

In the ASYNNNI model the internal degrees of freedom of the CFs
are ignored, such as the charge and spin distributions. To take them into
account, one may start from an appropriate  microscopic model of the
Cu-O chains and try to map it onto a
corresponding lattice-gas model, possibly of ASYNNNI-type.
(Microscopic models have been also employed by  Aligia 
\cite{alig_ssc,alig} to explain and predict some superstructures in the Cu(1) 
planes).
Such an approach has been attempted before \cite{yakh,gaw_2}, motivated
by the suggestion that, in the
case of YBCO, the ASYNNNI model may be insufficient to describe
phenomena in the part of the phase diagram where short Cu-O chains
dominate. For instance, the chemical potential isotherms \cite{schl} are
reproduced only poorly, whose singularities
seem to be important near the tetragonal-to-Ortho-I
phase boundary where
the average lenght of the chains does not exceed a few lattice constants.
It had been suggested \cite{alar,rey,hohl}
to describe the tetragonal phase in YBCO
at $x\!<\!0.4$ in terms of a phase separation. In particular, neutron and
x-ray data had been interpreted, in the regime of phase separation,
as signalling short chains with a herringbone structure. However,
subsequently, the superstructure reflections were attributed to a
parasitic phase \cite{yakh}. (Nevertheless, there may be a chance to
find herringbone structures
or some other regular configurations of the very short chain fragments
in LaBCO, but not in YBCO; for possible patterns, see Figure \ref{figure4}.
Indeed, it has been shown \cite{schreck} that those
chain fragments are electric quadrupoles, and if they are dominant,
then the herringbone structure would be energetically preferred).

The motivation of the present study is to reconsider the ASYNNNI model
in view of the new experimental facts.
We shall relate that lattice-gas model to a strongly correlated
electron model, the Emery model,
for the Cu-O chains. This will allow us to reinterpret the crucial
coupling $V_2$, which will turn out to depend, for instance, on
temperature. In addition, a new parameter will be introduced,
reflecting the relevance of the shortest chain fragments, i.e. Cu-O-Cu.
Using Monte Carlo simulations, properties of the resulting
{\it extended} ASYNNNI model will be computed, based on the statistical
ensemble of chain fragments. The main considerations in relating
the Emery model to the ASYNNNI model will be presented in the
next section, followed by a section in which, based on the lattice-gas
model, 
some interesting properties like the temperature dependence of
the charge transfer from the Cu-O chains in the Cu(1) planes to the
Cu(2) planes for various REBCO compounds as well as the distribution
of differently coordinated Cu atoms will be discussed.
A short summary will conclude the article.
\section{Microscopic description}
\subsection*{Charge transfer mechanism}
Following prior considerations \cite{rm}, we shall describe briefly
a possible charge transfer mechanism for the Cu-O chains in the Cu(1)
planes.

Assuming a strong repulsion between neighboring oxygen sites, see Figure
\ref{figure2}, linear Cu-O chain fragments are formed with essentially
no crossings. In a CF, oxygen and copper atoms form a strongly
correlated system
due to a strong hybridization of the oxygen $p_x(p_y)$ orbitals with the
$d_{z^2-x^2}(d_{z^2-y^2})$ orbitals of the copper atoms.
If we ignored the oxidation process, the $n$
oxygen atoms in a CF of composition Cu$_{n+1}$O$_n$ would be neutral,
while the $n\!+\!1$ Cu atoms would be in the monovalent
state, as in the parent compound YBa$_2$Cu$_3$O$_6$. In reality, a charge
transfer is expected to occur, which may be described formally as
a two-step process. The first
step corresponds to a charge redistribution in the CF, leading
to only one divalent O$^{2-}$ ion in each CF; schematically
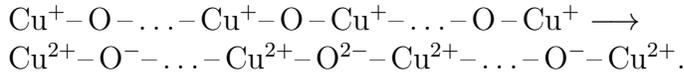
\newline
Cu$^{+}$--$\,$O$\,$--$\,\dots\,$--$\,$Cu$^{+}$--$\,$O$\,$--$\,$Cu$^{+}$--$\,
\dots\,$--$\,$O$\,$--$\,$Cu$^{+}\longrightarrow$
\newline
Cu$^{2+}$--$\,$O${^-}$--$\,\dots\,$--$\,$Cu$^{2+}$--$\,$O$^{2-}$--$\,
$Cu$^{2+}$--$\,\dots\,$--$\,$O$^{-}$--$\,$Cu$^{2+}.$
\newline
\noindent
The second step involves a charge transfer from the monovalent
oxygen ions in a chain fragment to the Cu(2) planes or to other fragments.
The resulting CF has the form
Cu$^{2+}_{n+1}$O$^{2-}_m$O$^-_{n-m}$, where $m-1$ holes have been
transferred, assuming that each copper ion remains in the divalent
state. It has been shown experimentally
\cite{krol} that for long chains in YBCO $\,m/n\!\approx\!0.7$.
For short chains,
the optimal CF configurations have been estimated \cite{rm} to be
Cu$^{2+}_5$O$^{2-}_3$O$^{-}$ for $n\!=\!4$,
Cu$^{2+}_4$O$^{2-}_2$O$^{-}$ for $n\!=\!3$ and
Cu$^{2+}_3$O$^{2-}$O$^{-}$ (or Cu$^{2+}_3$O$^{2-}_2$) for $n\!=\!2$. 
Accordingly,
no hole (one hole) leaves the $n\!=\!2$ CF, one hole and two holes
leave the $n\!=\!3$ and $n\!=\!4$ CF, respectively. Generalizing this
finding into a simple rule of thumb, one may estimate $m$ to be, for an
arbitrarily long chain,
~$m={\rm nint}(n(m/n)_{\rm opt})$, where nint denotes the nearest integer.
$(m/n)_{\rm opt}$ is that optimal ratio, which is realized in {\it
long} CFs.
\subsection*{The Emery model}
As the starting point for a microscopic description of the Cu-O chains
we use the one--dimensional Emery model \cite{emer}, applied to
systems of finite length.
The Hamiltonian is
\begin{eqnarray}
H=-t\sum_{<r,\rho>}\sum_{\sigma}\left(p^{\dagger}_{r\sigma}d_{\rho\sigma}+
d^{\dagger}_{\rho\sigma}p_{r\sigma}\right)+\sum_r\left(\epsilon_pn_r+
U_pn_{r\uparrow}n_{r\downarrow}\right)+\nonumber\\
+\sum_{\rho}\left(\epsilon_dn_{\rho}+U_dn_{\rho\uparrow}n_{\rho\downarrow}
\right)+U_{\rm end}\!\sum_{\widetilde\rho}n_{\widetilde\rho}
\label{1}
\end{eqnarray}
where the sum runs over the O sites ($r$) and the Cu sites ($\rho$) of a
{\it finite} chain, $<\!r,\rho\!>$ denotes neighboring lattice sites, and
$\widetilde\rho$ refers to the Cu$^{(3)}$ sites at the chain ends.
$d_{\sigma}$ and $p_{\sigma}$ ($d^{\dagger}_{\sigma}$ and
$p^{\dagger}_{\sigma}$) are annihilation (creation) operators of a hole with
spin $\sigma$ on Cu and O sites, respectively. $n_{\sigma}$ is the hole
(spin $\sigma$) occupation number and $n\!=\!\sum_{\sigma}n_{\sigma}$.
This model ignores the contribution
of the apical oxygen ions to the energy, which, indeed, play a minor
role, because they are essentially divalent (cf discussion in \cite{garc}).

Because of the quantum origin of alternating Cu-O chains, the hole
occupation number on Cu sites is not rigorously equal to 1, as tacitly
assumed when the charge transfer mechanism was discussed.
This occupation number fluctuates, remaining slightly below 1 ($\approx 0.85$,
as numerically estimated in \cite{gaw_1}). Now, a good quantum number
is the number of holes (both, oxygen and copper), say $\nu(n)$ in a CF
of length $n$. Nevertheless, $m$ remains a quantum number,
characterizing the hole transfer from a chain. The relation between
$m(n)$ and $\nu(n)$ is given by
\begin{eqnarray}
\label{n_m}
\nu(n)=2n-(m(n)-1).
\end{eqnarray}
There are strongly
correlated models, e.g., the Kondo-reduced Emery model (see  \cite{rm,haug}),
which allow to treat Cu ions of CFs as divalent, 
and oxygen ions as monovalent ($n-m$) or divalent ($m$).

Note that we include in model (\ref{1}) the shift
of the on-site hole energy level, $U_{\rm end}$, which is due to
non-equivalency of Cu$^{(3)}$ and Cu$^{(4)}$ ions. There may be alternate
ways to take into account
the $U_{\rm end}$-term; for instance Aligia uses
a model \cite{alig_ssc,alig} which includes the
interaction of charges of neighboring atoms, $\rho$(Cu) and $r$(O),
\begin{eqnarray}
U_{\rm pd}\sum_{<r,\rho>}(1+n_{\rho})(-2+n_r)
\nonumber
\end{eqnarray}
Our description could be interpreted as a simplified version of the
$U_{\rm pd}\,$-term, $U_{\rm end}\approx U_{\rm pd}(2-\overline{n_r})$,
where $\overline{n_r}$ denotes the thermal expectation value for the
occupancy of an oxygen site.

In the case of RE=Nd or La, the ionic radii of the rare-earth ions
are larger than in the other REBCO compounds. Consequently,
the interatomic distance is also larger, but the hole hopping
amplitude is smaller -- in the chains, and in the planes. We shall
discuss below why this circumstance may favor short
chain fragments, depending on
the oxygen hole chemical potential, $\mu$,
which varies in the REBCO series.
\subsection*{Free energy}
We consider the Cu$^{2+}_{n+1}$O$^{2-}_m$O$^-_{n-m}$ chain fragment,
described by the Emery model (\ref{1}). We may determine, say, numerically,
its free energy (here and
in the following, the free energies are taken per oxygen atom),
$f(n;m)$. Indeed,
performing statistical averaging with respect to the internal, fermionic,
degrees of freedom ($m$ and possible spin arrangements), and disregarding
the configurational entropy, the free energy for long chains ($n\gg 1$) has
been recently shown to have the form \cite {uim_pr}
\begin{eqnarray}
\label{0-3}
\phi(n)\approx\phi_1-
\frac {\phi_2}n -\frac T{2n}\ln n \,.
\end{eqnarray}
where $\phi_1$ and $\phi_2$ are coefficients which are given by
the energy parameters of the Emery model; $\phi_2$ is, approximately,
linear in $T$.

In specifying the parameters
of the Emery model, we largely follow  Refs.\cite{gaw_1,gaw_2}.
$\epsilon_d$ is taken as the reference level.
Measuring the other energy parameters in  units of $t$, we set $\epsilon_p=2$,
$U_p=4$ and $U_d=8$, similar to the values in the CuO$_2$
planes of YBCO. The REBCO compounds are expected to differ mainly in the
value of the hopping amplitude $t$ (being approximately
$t=1.35$ eV $\approx 15000$ K for YBCO).
As noted before \cite{gaw_1,gaw_2}, the remaining two parameters,
the chemical potential $\mu$ and $U_{\rm end}$, play an essential role
in favoring long or short chains. $\mu$ equilibrates the hole exchange
between the two oxygen hole reservoirs, CuO$_2$ planes and CFs.
$U_{\rm end}$ stems from the non-equivalency of the Cu$^{(3)}$ and Cu$^{(4)}$
sites ($\epsilon_d+U_{\rm end}$ denotes the hole level on a Cu$^{(3)}$-site).
Although $U_{\rm end}$ is positive, as seen from its relation to
$U_{\rm pd}$, its exact value is not known,
neither from experiments nor from band-structure calculations.

In YBCO the oxygen atoms of the oxygen deficient planes tend to
arrange themselves into long chains at low enough quenching temperatures;
probably even at quite low oxygen content. In contrast, in
LaBCO long chains are predominant only at $x\sim 1$ \cite{luetg2}.
In the REBCO compounds, the various tendencies in the chain lenghts
are expected to be caused by the various ionic radii of
their rare-earth constituents (see Table \ref{IONICRADII}).
In the model description (\ref{1}) of those compounds, the crucial
parameters are the hopping amplitude $t$ which depends on the RE ionic
radius, and the chemical potential which regulates
the degree of filling of the 1D O-hole band.

To see whether short or long Cu-O chains are favored, we
compute the free energy per oxygen atom as a function of the
chain length.
If it increases with length, then short chains are more likely,
otherwise longer chains are preferred.

For a given set of parameters one may calculate the free energy in
the grand canonical
ensemble, thus incorporating the chemical potential $\mu$, which
governs the concentration of holes in a chain. Following this scheme
one needs, in principle, the whole set of energy levels of the
Emery model. However, since the temperatures we are dealing with are
much lower than
the typical electronic energies, which are of order several
thousand K, it is sufficient
to include the lowest energy levels in the free energy
calculations. These levels may be determined by using
the Lanczos algorithm \cite{lan}, as we did in evaluating the free energy
of chains with up to $n=6$ oxygen atoms (plus 7 copper atoms).
The detailed procedure for calculating
the free energies is described in Appendix A.
\section{The extended ASYNNNI model}
\subsection*{Estimate of effective parameters}
The free energy of the chains described by the Emery model,
expression (\ref{0-3}), corresponds in the standard ASYNNNI model to the
energy of a Cu-O chain with $n$ oxygen atoms, neglecting
interchain interactions, i.e.
\begin{eqnarray}
\label{0-4}
\widetilde\phi(n)=V_2-\frac {V_2}n \,.
\end{eqnarray}
A clear difference of $\phi(n)$ and $\widetilde\phi(n)$ is that
$\widetilde\phi(n)$ decreases monotonically with $n$ at $V_2<0$,
whereas, if the analogous condition $\phi_2<0$ is satisfied,
$\phi(n)$ exhibits a minimum at finite, but exponentially large value
$n_{\rm opt}\propto\exp 2\phi_2/T_q$. It is estimated
as exceeding a hundred 
at temperatures below 300$^\circ$ C. Hence, the logarithmic
correction in the r.h.s. of Eq.(\ref{0-3}) may start playing an important
role only at $x\sim 1$.

$\phi_2$ in Eq.(\ref{0-3}) depends on temperature, while $V_2$ does
not (in both cases, we did not consider the configurational entropy).
Accordingly, accepting the Emery model as
the more fundamental description for the chain fragments, the
ASYNNNI model may be modified by introducing an effective,
{\it temperature--dependent} coupling $V_2$,
corresponding to $\phi_2$ in the Emery
model. The difference in the coefficients $\phi_1$ and $\phi_2$ may not break
the analogy with the ASYNNNI model. In relevant experiments, the CF ensemble
is measured at a fixed oxygen content, thus $\phi_1$ contributes 
equally to any realization of CFs and may be rescaled,
say, to the value of $\phi_2$.

Note that Eq.(\ref{0-3}) holds only for sufficiently
long chains. Indeed, for short chain fragments,
$\phi(n)$ may show oscillations \cite{gaw_2}. To include these subtle
features in a statistical description (for example, in a 
Monte Carlo simulation),
one may deal with the free energies of the short chains
separately. The largest deviation from the expressions (3) and (4)
occurs at $n=1$. Thence, to incorporate those oscillations
in the simplest fashion,
we consider an {\it extended} ASYNNNI model, attributing to the
chains with $n=1$, Cu-O-Cu, an additional 
energy $\widetilde{V}_2$. For larger values of $n$, we keep the parameters of
the standard ASYNNNI model. Accordingly, the free energy
may be cast in the form
\begin{eqnarray}
\label{fn}
\widetilde\phi(n)=V_2-\frac {V_2}n, \quad n\!>\!1; \qquad
\widetilde\phi(1)=\widetilde{V}_2 \,.
\end{eqnarray}

To substantiate Eqs.(\ref{0-3}) and (\ref{fn}), we shall
now present results of calculations of the free energy for the
Emery model of Cu-O chain fragments.

Let us first roughly estimate a realistic value of $U_{\rm end}$.
In Figure \ref{Uend}, the free energy of the Emery model is shown,
using units appropriate for YBCO (we have chosen 
the chemical potential $\mu$ at each $U_{\rm end}$ to assure that
the concentration of oxygen holes in the {\it longest} chains we
studied is
$\approx$ 0.3).
Specifically, the energy parameters are taken in units of
$t = 1.35$ eV $\approx 15000$ K, so that the temperature range of
experimental interest ($T_q$ between 300 K and 1200 K) is between 0.02
and 0.08. Comparing the free energies for long chains
to expression Eq. (\ref{0-4}) (disregarding, at this stage, $\widetilde V_2$)
one finds that $U_{\rm end}=1$ corresponds to a moderately attractive
$V_2$, of several hundreds of Kelvin in agreement with the experimentally
observed disordering of the Ortho-II phase at about 450 K. Otherwise,
for example, for $U_{\rm end}=0$ the free energy grows too rapidly with
the chain length, corresponding to an effective repulsion $V_2$ of the order
of a few thousands of Kelvin (see also \cite{alig}). Increasing $U_{\rm end}$
leads first to a weaker repulsion,
and finally an attractive $V_2$.
For instance, $U_{\rm end}=1.5$ would imply a value for $V_2$ of the order of
several thousand K, being a strong effective attraction. Thereby,
by tuning $U_{\rm end}$, one may stabilize either short chain fragments
(some realizations at $x\leq 1/2$ are shown in Figure \ref{figure4})
or very long Cu-O chains.

While $U_{\rm end}$ is expected to be essentially a constant for
all REBCO compounds, 
the chemical potential $\mu$ may vary, e.g., with the
oxygen content $x$ or the RE element.
In Figure \ref{mu} the free energy per oxygen is shown for several values
of the chemical potential at the particular choice $U_{\rm end}=1$.
With $\mu$ increasing, the effective intra-chain interaction $V_2$
between the oxygen atoms in the Cu-O chains
changes from attraction to repulsion. This is happening in a fairly broad
range, $\mu$=1.5 - 2.0, where
$V_2$ goes from a moderately attractive value ($\sim -0.15$) to
a moderately repulsive one ($\sim 0.1$). Changing the RE element in the
REBCO compounds implies changing $\mu$ and hence the intra-chain interaction.
As will be discussed later, $V_2$ varies slowly, remaining $\sim -0.1$
in the range of relevant values of $\mu$. In fact,
$\widetilde V_2$ introduces an important competing effect.

In Figure \ref{FreeFit} we depict the free energy per oxygen atom versus $n$
for $U_{\rm end}= 1$ at different temperatures, choosing the
chemical potential so that the concentration of oxygen holes is, for long
chains, approximately 0.3, as it is the case for YBCO. 
$\mu \approx 1.6$ is well compatible with that concentration.
One may easily see that the lower the (quenching) temperature $T$
the larger the effective intra-chain oxygen-oxygen attraction will be.
In addition, at low $T$, the free energy shows an oscillatory
behavior for small values of $n$ (see, for example, Figure \ref{FreeFit}
at $T= 0.02$). Such oscillations are maximal at $T=0$
(see also Figure 1 in \cite{gaw_2}).
It is obvious from Figure \ref{FreeFit} that the fitting of the free
energy to that of the standard ASYNNNI model, Eq.(\ref{0-4}), is
unsatisfactory even in the high temperature region.
The $n=1$ case has to be fitted separately, defining $\widetilde V_2$ in
the extended ASYNNNI model, Eq.(\ref{fn}).
It should be emphasized, that
the effective parameters of the hereby defined {\it extended} ASYNNNI
model, $V_2$ and $\widetilde{V}_2$, depend not only on the temperature,
but also on $U_{\rm end}$. In addition, they depend on a specific
filling of the 1D oxygen hole band, i.e. $\mu$.

The other two couplings of the full ASYNNNI model, $V_1$
and $V_3$, are
of predominantly electrostatic origin, and may be supposed to be
constant. Of course, their values do not follow
from the Emery model. (At this
point, attention may be drawn to another attempt
to determine
parameters of the (standard) ASYNNNI model from first-principles,
based, in that case, on linear muffin-tin-orbital calcualtions for
YBCO \cite{STERNE89}. In that analysis, only ground state properties
have been considered, leading to constant couplings $V_1$, $V_2$, and
$V_3$.)

In Figure \ref{FreeFit}, also least-square fits of the free energy for
the Emery model to the ansatz, Eq. (5),
defining the extended ASYNNNI model are
shown, at several temperatures. From such fits one may
determine $V_2$ and $\widetilde{V}_2$
as function of temperature. Examples are shown in Figures \ref{FreeFitV2}
and \ref{FreeFitV2tilde} for a few values of $U_{\rm end}$ around
$U_{\rm end}=1$, presuming a fixed hole concentration of 0.3 in the longest
chains.
It is evident from Figure \ref{FreeFitV2} that $V_2$ displays a pronounced
linear dependence on $T$ (cf, Eq.(\ref{0-3}), where $\phi_2$ is also linear in
$T$), and  $V_2$ is practically linear in $U_{\rm end}$ at a given temperature.

To fix $U_{\rm end}$, one may compare results on the extended ASYNNNI model
to experiments. A possible procedure will be demonstrated now for YBCO.
There we estimate $U_{\rm end}$ by locating the disordering
transition of the Ortho-II phase
of YBa$_2$Cu$_3$O$_{\rm 6+x}$ with
$x = 0.5$,  known to occur experimentally at
about 450 K \cite{gant,veal_3,OIITc,schw}. That
transition temperature is found to depend rather sensitively on the
choice of $U_{\rm end}$, and hence $V_2$ and $\widetilde{V}_2$ in the
framework of the extended ASYNNNI model. The
lattice gas model is analysed by
using standard Monte Carlo (MC) techniques.

To simulate the extended ASYNNNI model, one first has to specify the
couplings $V_1$ and $V_3$. Different values
have been proposed and studied for the conventional ASYNNNI model
\cite{font_1,selk,OLDSET2}. Here we adopt values which have been
argued to reproduce the experimental structural phase diagram of
YBCO fairly well \cite{fiig1,fiig2}, namely, in terms of $V_2$,
\begin{eqnarray}
V_1 = -2.86\cdot V_2 \label{Newset1}\\
V_3 = -0.46\cdot V_2 \label{Newset2}
\end{eqnarray}
As discussed above, $V_2$ is expected to depend on temperature
much more strongly than $V_1$ and $V_3$. Thence, in fixing
$V_1$ and $V_3$, we choose the value of $V_2$ at the Ortho-II disordering 
temperature, $T_{\rm II}\approx 0.03$ (note that we checked, by
simulations, that choosing the room temperature $T_R$, instead of $T_{\rm II}$,
leads to rather minor changes of $V_1$ and $V_3$, of a few percent, 
and affects the estimate for $U_{\rm end}$ only mildly).

Simulations at $T_{\rm II}$ for different values of 
$U_{\rm end}$ (and tuning $\mu$ to reproduce the hole concentration 
for the longest chains), which determine
$V_2(T_{\rm II};U_{\rm end})$ and 
$\widetilde{V}_2(T_{\rm II};U_{\rm end})$
have been performed. The order 
parameter $m_{\rm \footnotesize II}$ \cite{selk,fiig2} of the Ortho-II phase
is found to drop
abruptly at $U_{\rm end}=1.02$,
see Figure \ref{3DOII}. From Figures \ref{FreeFitV2} and \ref{FreeFitV2tilde},
one obtains $V_2(T_{\rm II})\approx -0.12$ and $\widetilde{V}_2(T_{\rm II})
\approx -0.09$. Summarizing, we have finally estimated, for
YBCO, $U_{\rm end} = 1.02$ and $\mu =1.59$
from the experimental values of $T_{\rm II}$ and
the hole concentration for long chains. 

In substituting yttrium by another RE element, the chemical potential
will be changed. This, in turn, will lead to different values of the
effective
parameters in the ASYNNNI model. Results of pertinent calculations
are depicted in Figure \ref{MUVARIV2}, taking $U_{\rm end} = 1.02$.
By changing $\mu$ in between 1.55 and 1.70, one induces only
mild changes in $V_2$, in between -0.15 and -0.1. More importantly,
we observe in that range a crossover in the tendency to form short or
long Cu-O chains. Obviously, that tendency follows from the
fact whether $\widetilde{V}_2$ is smaller than $V_2$, or vice versa.
At $\mu =1.59$ (characteristic for YBCO), the long chains are favored;
at $\mu$ $\approx 1.65$, a compensation holds. At larger values of the
chemical potential, short chains will be preferred.

Accordingly, for each REBCO compound, one may predict the preference
for short or long chains from the value
of $\mu$. Although the chemical potential is not
precisely known for the various RE cases, one may safely assume that
it increases with the ionic radius. As shown in
table \ref{IONICRADII}, these radii vary appreciably. For instance, they
are larger for Nd and La than for Y.

A microscopic description for estimating $\mu$ should invoke the
structure of the oxygen hole bands of the chain fragments (1D) as well
as of the Cu(2) planes (2D). It has been assumed in \cite{gaw_2}
that the 1D oxygen hole band is 
lower than the 2D band, with an overlap,  
as shown schematically
in Figure \ref{figure_12}, where $\mu$ is above the bottom of the 2D 
hole band.
Despite a potentially significant charge transfer from
chains (i.e. $\overline{m(n)}-1$ from each CF of length $n$),
such a band structure is no contradiction. In fact, at small
oxygen content $x$, shorter 
CFs are preferred, the charge transfer gets effectively surpressed.
The holes transferred from chains become partly localized on the
impurity levels and partly
2D carriers, with a rather low, semiconducting, concentration.
In the metallic state the chemical potential intersects the
2D band; a sufficiently large hole transfer from chains to planes causes
an electrostatic shift of the levels, and the 1D band raises with
respect to the 2D band. Based on such a description,
the transformation from the semiconducting to the metallic state, by
changing the oxygen content $x$, can be qualitatively described
\cite{gaw_2,comment}.  

In principle, one could estimate the variation of $\mu$ with $x$ from
a self-consistent scheme which would be very tedious, based upon
experimental data which have not been established reliably, so far. In
the following analysis, we shall, however, assume that the
chemical potential is independent of $x$. This will be sufficient
to reproduce the main trends resulting from the change in
$\mu$ in the REBCO series.
To describe the corresponding thermal properties
of the chain fragments, we shall investigate the extended ASYNNNI
model, with the parameters $\widetilde{V}_2$ and $V_2$ changing with
the chemical potential. For $V_1$ and $V_3$, we shall always take
the values for YBCO, see above. Because they are associated with
Coulomb forces on small distances, they are expected to vary only
slightly, by a few percents, when substituting Y by, say, Nd or La.
As we checked in simulations, such a variation will not affect the
conclusions. The results of the Monte Carlo study will be compared
to experimental findings for LaBCO and NdBCO \cite{luetg1,luetg2}.
\subsection*{Monte Carlo simulations, Comparison with experiments,
Discussions}

The simulations of the extended ASYNNNI model describing REBCO
compounds have been performed for systems with L $\times$ L Cu sites,
imposing full periodic boundary conditions. Usually,
L was taken to be 40; to estimate the role of finite size effects,
a few runs were done with larger lattices. We employed
Kawasaki dynamics, keeping the oxygen content $x$
constant during a simulation, and choosing the initial and final site
of an oxygen atom randomly. The new site, if empty, is accepted with a
probability given by the Boltzmann factor of the change in energy
of that possible move.
An updating scheme which restricts the hopping of the oxygen atoms to
neighboring lattice sites would be more realistic in mimicing dynamical
processes, but slower. Because we are here only interested in
equilibrium properties, we employed the faster algorithm.

By varying the oxygen content $x$ and the temperature $T$, the
structural phase diagram may be determined.
Figure \ref{PHASEDIAGRAM} depicts the phase diagram for the case
of YBCO, using the values of the effective parameters discussed above.
The phase boundaries have been estimated from
the behavior of the order parameters of the various phases and
the positions of the maxima in the specific heat, as before \cite{selk}.
The topology of the phase diagram is similar to that obtained for
the standard ASYNNNI model in prior computations, e.g. \cite{selk,fiig1},
including the Ortho-I, Ortho-II, and
tetragonal phases. Perhaps most notably, the maximal phase transition
temperature of the Ortho-II phase is, however, somewhat reduced, in
accordance with the experimental findings.

By analysing Monte Carlo equilibrium configurations one may
easily identify the Cu-O chain fragments and hence
determine, for instance, the thermally averaged concentration of
differently coordinated Cu ions.
In Figure \ref{CU2} the concentration of Cu$^{(2)}$ ions, $c_2$, is shown as a
function of the oxygen content
for various temperatures, in the YBCO case. While at high temperatures
$c_2$ decreases monotonically with increasing oxygen content, one
finds a non-monotonic or plateau-like behavior when lowering the temperature,
so that one may encounter the Ortho-II phase. Obviously, in the perfect
Ortho-II structure, at $x=0.5$ and sufficiently low temperatures,
half of the Cu atoms belong to the perfect Cu-O chains extending
throughout the lattice and half of them are in oxygen-free rows, see
Figure \ref{figure1}(c). Accordingly, the concentration
of Cu$^{(2)}$ ions in such an ideal configuration is $0.5$.
Experimental results of XAS measurements at low and high
temperatures
\cite{tolen} are included in Figure \ref{CU2}. Agreement
between experimental data and theoretical curves is quite satisfactory,
except for
the point at $x\approx 0.39$, low $T_q$, which may, indeed, have been
determined inaccurately in the experiment. The behavior of $c_2$ versus $x$
at $T= 450$ K also agrees rather well with the experimental results 
using NMR-NQR \cite{luetg1,luetg2}. The concentrations of
Cu$^{(2)}$, Cu$^{(3)}$ and Cu$^{(4)}$ ($c_2$, $c_3$, and $c_4$) obtained
from the
MC simulations at $T = 450$ K in the case of YBCO are
shown in Figure \ref{CUfoldY}, together with experimental data
(their accuracy has not been mentioned) \cite{luetg2}.
A few experimental points deviate clearly 
from the theoretical curves.
Note, that the experimental data of \cite{luetg2} plotted in Figure
\ref{CUfoldY} obey well the sum rule
\begin{equation}
\label{rule1}
c_2+c_3+c_4=1
\end{equation}
This identity assumes that all Cu ions in the Cu(1)
planes can be identified as Cu$^{(2)}$, Cu$^{(3)}$
or Cu$^{(4)}$. With that assumption, a second sum rule is evident
\begin{equation}
\label{rule2}
c_3+2c_4=2x.
\end{equation}
These two sum rules allow to express two of the concentrations, say,
$c_3$ and $c_4$, by the remaining one, $c_2$, and the oxygen content $x$;
for instance,
\begin{eqnarray}
c_3=2-2x-2c_2, \quad c_4=2x-1+c_2.
\nonumber
\end{eqnarray}
The concentrations are allowed to vary in the range defined by the following
inequalities,
\begin{eqnarray}
\label{ineq}
\begin{array}{r}
0\leq x\leq 1/2 \longrightarrow 1-2x\leq c_2\leq 1-x, \quad 0\leq c_3 \leq 2x,
\quad 0\leq c_4 \leq x;\\
1/2\leq x\leq 1 \longrightarrow 0\leq c_2\leq 1-x, \quad 0\leq c_3 \leq 2-2x,
\quad 2x-1\leq c_4 \leq x.
\end{array}
\end{eqnarray}
A few points of the experimental set disagree significantly with
(\ref{rule2}), several points are beyond the limits defined by (\ref{ineq}).
Note that there is an uncertainty, reaching up to 0.1 according to
\cite{luetg2}, in determining the
oxygen content in the range $0.3<\!x<\!0.4$ (probably, in
the region of phase separation).

As mentioned above two concentrations can be determined by the
remaining third concentration.
Taking the latter one as $c_2$ for $x<\!0.5$, $c_3$ for $0.5\leq x<\!0.7$,
$c_4$ for $0.7<\! x\leq 0.8$, and again $c_2$ for $0.8<\!x$ ($c_4$
would be beyond
the limit of the inequality), we can fit nicely the revised set of
experimental points to theoretical curves in Figure \ref{CUfoldY}.

Similar measurements on the concentrations of differently coordinated
Cu ions have been reported for
NdBa$_2$Cu$_3$O$_{6+{\rm x}}$ (NdBCO)
and LaBa$_2$Cu$_3$O$_{6+{\rm x}}$ (LaBCO) \cite{luetg2}. As argued above,
we propose that the effect of the increase of the ionic radius
in Nd and La compounds can be incorporated into the microscopic model by
an increase in the chemical potential, leading to modified effective
parameters in the extended ASYNNNI model. Although $\mu$
is expected to depend on the oxygen content $x$ for each
REBCO compound, we consider a constant $\mu$ for a given RE, for reasons
stated before. This simplification seems to be justified in order to
detect the main trends.
Indeed, when increasing $\mu$ from 1.59, for YBCO, to
1.62 and 1.66, for NdBCO and LaBCO, respectively, the simulated data on
$c_2$, $c_3$, and $c_4$ agree rather well with the experimental results,
see figures \ref{CuFoldNd} and \ref{CuFoldLa}. For NdBCO we can use $c_3$
for determining the other concentrations in the range $0.5<\!x<\!0.7$,
as in the YBCO-case. For LaBCO, $c_3$ and $c_4$ have not been accessible
for direct measurements in the range $0.5<x$; in	
this case $c_2$ plays the determining role.

There may be also five- and sixfold coordinated Cu
ions in Cu(1) planes, due to a crossing of perpendicular chain
fragments. For example, Cu$^{(5)}$ would appear, if an oxygen atom
formed a 90$^{\circ}$ configuration with two other oxygen atoms (all three
are supposed to be adjacent to the same Cu atom). The energy loss
due to a  Cu$^{(5)}$ (Cu$^{(6)}$) ion would be $2V_1$
($4V_1$). From the simulations, the concentration of such ions
is seen to be very small for all oxygen contents $x$. A kink in
a CF would be energetically somewhat more favorable, albeit being
still rather costly, $V_1$. At the kink, a fourfold coordinated Cu
ion occurs, although it differs from the conventional Cu$^{(4)}$ ion.
Such a kink configuration cannot easily be described
by the Emery model.

Another important quantity related to the chain
fragments of the oxygen deficient Cu(1) planes, is the charge (hole) transfer,
$n_h$, to the CuO$_2$ planes.
It depends on the chemical potential as well as on other external parameters
such as $x$ and $T_q$.
$\overline{\nu_h(n)}$, the expectation number of holes (thermal averaging) 
leaving a CF of length $n$,
should be $\overline{m(n)}-1$ (see Figure \ref{ch_tr}, showing
a rather non-trivial dependence on $n$); 
it is associated with the expectation number of
oxygen and copper holes $\overline{\nu(n)}$ through (\ref{n_m}).
For long CFs in YBCO, one has $\overline{m(n)}/n\approx 0.7$; but for
short chain fragments, it may differ significantly.
The average concentration of oxygen holes
transferred from the chains, the charge transfer $n_h$, may be
approximated as, see \cite{selk},
\begin{eqnarray}\label{nh}
n_h(\mu;x;T_q)=\frac 1{L^2}\sum_{n=1}^L\frac{\overline{m(n)}-1}{2}
<\!N_{\small \rm CF}(n)\!>,
\end{eqnarray}
where $<\!N_{\small \rm CF}(n)\!>$ is the average number of CFs of
length $n$. The factor of 1/2 appears, because
holes from one Cu(1) plane are supplied to two Cu(2) planes.

Note that one can calculate the number of holes $\nu(n)$
in the framework of
the Emery model, at least for fairly short chains, see Appendix A,
which may then be extrapolated to longer chains.

Figure \ref{CHARGETRANSFER} shows the charge transfer
versus oxygen content for a set of equidistant temperatures,
obtained from simulating the extended ASYNNNI model for the YBCO
case. One observes that below $x\approx 0.25$
the charge transfer to CuO$_2$ planes is extremely small. In this regime,
in the tetragonal phase,
short chain fragments dominate, contributing much less to the
charge transfer than the long chains. The preference for short chains
could be reduced by going to very low (quenching) temperatures which,
however, are usually not accessible experimentally (recall 
that $T_q$ is bounded
by $T_R$ from below).
The tetragonal phase is still semiconducting, and the holes
are mainly localized in CuO$_2$ planes in the vicinity of the CF from
which they originated. The formation of longer CFs leads to a
significant increase of the
hole transfer, giving rise to the
semiconducting-to-metal transformation.

The charge transfer displays a plateau-like behavior
near $x = 0.5$, reflecting the well-known 60 K plateau in the
transition temperature to the superconducting phase in YBCO
at that oxygen content. This aspect has been discussed in quite
a few publications before.
In many experiments the close relationship between the superconducting
transition temperature $T_c$ and $n_h$ has been found
\cite{TORR89,SHAFER89}.
For instance, $T_c$ was argued to be linear \cite{BHB92,Rotter91}
or quadratic in $n_h$ \cite{Whangbo91,VEAL91}.
In both cases, the plateau in $T_c$ would correspond to one in $n_h$.
A second,
so-called 90 K plateau in $T_c$ versus $x$ occurs at $x > 0.8$, where
$n_h$ increases monotonically, while $T_c$ remains almost
constant. Overdoping is believed to be responsible for
that plateau \cite{lant_1,Whangbo91}: $T_c$ practically saturates when
$n_h$ becomes larger than the optimal hole concentration ($\approx 0.2-0.25$).

In Figure \ref{CHARGETRANSFERall} the charge tranfer $n_h$ versus oxygen
content $x$ is depicted at $T\!=\!450$ K for three values of the chemical
potential, mimicing, as before, YBCO, NdBCO, and LaBCO, i.e. $\mu$
= 1.59, 1.62 and 1.66. One readily sees, in accordance with
the experiments, that
the low-temperature plateau in $T_c$ versus $x$ is expected to be
much less pronounced in NdBCO \cite{veal_4,BHHKW90} and absent in LaBCO
\cite{lindemer}.
In that figure, we also included Monte Carlo data for $\mu = 1.56$, mimicing
a REBCO compound where the rare earth ions would be even smaller
than yttrium, like YbBCO (see table \ref{IONICRADII}). Indeed, it was
shown experimentally \cite{BHHKW90}
that the 60 K plateau in YbBCO is even more
manifest than in YBCO, in accordance with the MC data.

MC simulations allow to compute easily the average length of
Cu-O chain fragments, $\ell$.  Results are displayed in Figure
\ref{CHAINLENGTH}. $\ell$
is given by the ratio (total length of CFs)/(total number of CFs).
The numerator is identical with the number of oxygen ions, while the
denominator is, neglecting kinks and crossings of CFs,
one half of the Cu$^{(3)}$ atoms. Thence, one would obtain
$\ell=2x/c_3$. 
That formula has been used in \cite{luetg2}, to express
the experimental findings for $c_3$ for YBCO, NdBCO, and LaBCO in terms of
$\ell$. However, care is needed in doing that. In
Figure \ref{CHAINLENGTH}, we compare Monte Carlo results for $\ell$ invoking
the approximate formula to those using the correct expression for the
average chain lenght. Indeed, the overall agreement is fair, but 
there are obvious differences near $x\sim 1$.
\section{Summary}
The present paper has been motivated, {\it first}, by recent experimental
data on Cu-O chain fragments
\cite{luetg2}, and, {\it second}, by the wish of elaborating and
concretizing the idea to relate a microscopic description of those
fragments to a statistical analysis, i.e. to relate the quantum mechanics of
individual chains to the statistical mechanics of their ensembles
\cite{gaw_2}. Our main considerations and findings may be summarized
as follows.
\begin{itemize}
\item 
We consider the Emery model, which belongs to the family of strongly 
correlated fermionic (hole) models, incorporating into it the effect of chain
ends ($U_{\rm end}$-term). We assume that the lowest, Cu-like, band is 
completely filled, whereas the intermediate, O-like, band is partly occupied.
The occupancy of the latter band is regulated by the chemical
potential $\mu$, which determines the equilibrium of the
oxygen hole reservoirs in the CuO$_2$ planes and CFs.
\item 
In accordance with experimental results, the
chemical potential level is assumed to be located in the 1D oxygen hole
band of the Cu-O chain fragments. The oxygen hole concentration
in long CFs is determined by the chemical potential $\mu$.
\item 
The oxygen hole number in a CF follows from the number of
holes leaving the chain. There is essentially no charge transfer from short CFs
of lengths $n=1$ and 2. This explains the predominance of fairly short CFs 
in the semiconducting state of the REBCO compounds.
\item
The "semiconductor-metal" transformation is governed by the dependence of the 
chemical potential on the oxygen content $x$.
Analysis of finite chains by means of the Emery model shows that CFs
of minimal length occur less often when $\mu$ decreases.
This is expected to be accompanied by an increasing number of holes
transferred from chains to planes.
Coulomb forces lead to the lowering of $\mu$, which favors
longer CFs, giving rise to an additional charge transfer, etc.
\item
In our analysis, the REBCO compounds have been characterised by different, but
fixed values of $\mu$. However, in reality, with
$x$ increasing, $\mu$ is believed to decrease
proportionally to $n_h$, i.e.
the hole doping of the CuO$_2$ planes. Using Figure \ref{MUVARIV2},
these changes of $\mu$ could be incorporated into the calculations.
\item
An important role in the Emery model is played by $U_{\rm end}$;
the energy loss stemming from the ends of a CF determines the effective
attraction or repulsion between the oxygen atoms in CFs.
In fact, ignoring this term would result in
a very strong O-O repulsion in the chains (of the order of several
hundreds meV in the case of YBCO, where
the concentration of oxygen holes in long CFs is about 0.3).
Experimental data allow one to estimate $U_{\rm end}$ to be
$\approx 1$ in units of the hopping term $t$ of the Emery model.
\item
The free energy of the Emery model has been calculated
for individual CFs, characterized by their length
$n$, as a function of the chemical potential $\mu$,
$U_{\rm end}$, and the temperature $T$. For specific REBCO compounds,
$\mu$ and $U_{\rm end}$ can be fixed using 
experimental data, e.g., for the oxygen hole concentration and the disordering
temperature of the Ortho--II phase. By comparison of the free energy,
the values of the parameters of an {\it extended} ASYNNNI
model can then be
estimated. In the framework of that model, the structural phase
diagrams of the REBCO compounds can be conveniently studied, using,
for instance, Monte Carlo simulations.
\item
The larger the ionic radius of the REBCO compound the higher the
chemical potential, and the shorter the Cu-O chain fragments.
\item
Analysing the extended ASYNNNI model for various RE constituents,
we also computed macroscopic properties 
such as the concentration of differently coordinated Cu ions
and the charge transfer
from the chain fragments.
\end{itemize}

In conclusion, the theoretical approach presented in this paper
(start from microscopic calculations, identify effective
parameters of a statistical model as well as, based on
experimental results, their values, and finally perform simulations),
allows to reproduce main trends in the statistical properties
of Cu-O chain fragments
throughout the various REBCO compounds. Refinements and extensions,
based on additional experimental data, are possible.
\subsection*{Acknowledgement}
We should like to thank H. L\"utgemeier for informing us about his
experimental results. Financial support by the Russian Foundation for Basic
Research, the Research Council of Norway (Programme for Supercomputing), 
and by INTAS (Project 93-211) is gratefully
acknowlegded.
\appendix
\section{the free energy calculation for chain fragments}
The Hamiltonian of the Emery model for a CF, characterized
by $n$ and $m$, deals
with $\nu=2n+1-m$ fermions which are spin-1/2 holes. Positive
integers $m$ ($>1$)
describe the charge transfer from a CF.
The eigenstates of this Hamiltonian can be easily divided into
blocks which correspond to different
numbers $\nu$ ($\nu=0,1,\dots,2(2n+1)$) and, at
given $\nu$, to different total spin projections
($S_z = -\nu/2, -\nu/2+1,\dots, \nu/2$). Any block
can be further subdivided into two blocks which consist of either symmetric or
antisymmetric states with respect to the central Cu atom of a CF.
The size of actual blocks grows fastly with $n$: CFs with $n = 6$ oxygen
atoms, plus 7 copper atoms, are still
accessible for calculations of a substantial number of the low lying energy
levels.

Our conjecture is that the CF properties, including the concentration of
holes, depend on
the chemical potential $\mu$ which, in turn, is a function of the hole
doping. Thus, we must calculate the grand canonical partition function:
\begin{equation}
\Xi_n(T,\mu) = \sum_{\nu = 0}^{2(2n+1)}\sum_{S_z=-\nu/2}^{\nu/2} \sum_{i}
		e^{-\left[E_i(\nu,Sz) -\mu \nu\right]/T},
\end{equation}
where $i$ runs over all the states within the block at
given $\nu$ and $S_z$. The temperatures of interest do not exceed
1200 K, which is low on the energy scale of the model (\ref{1}). Indeed,
e.g., including one thousand
low lying energy levels at $n=6$,
we can properly evaluate
the partition function (doubling the number of low lying states, one
does not change the free energy significantly).

According to \cite{krol}, one has
$\overline{m}/n\approx 0.7$, provided the copper hole occupation number
is equal to one. We
incorporate this fact, demanding the average total hole number
$\overline{\nu}=7+6\cdot 0.3\approx 8.8$ for the longest chains ($n=6$
oxygen and seven copper atoms),
analysed using
the Lanczos algorithm \cite{lan}. Thus, $\overline{\nu}$ is
defined as
\begin{eqnarray}
\overline{\nu} = \frac{1}{\Xi_n}\sum_{\nu = 0}^{2(2n+1)}
\sum_{S_z=-\nu/2}^{\nu/2} \sum_{i} \nu
		e^{-\left[E_i(\nu,Sz) -\mu \nu\right]/kT}
\end{eqnarray}
and $\mu(T)$ is then numerically determined from the constraint
$\overline{\nu}= 8.8$ for $n=6$.

In the framework of the Lanczos algorithm \cite{lan}, applied to a block of
the states $\{\nu,S_z\}$, we can find, say, the $M$ lowest eigenvalues.
$4M$ iterations should guarantee convergence of $M$ eigenfunctions.
There may be a problem with the Lanczos algorithm:
it fails to distinguish two or
more degenerate eigenvalues. However,
this is not a serious problem in our
case, since the accidental degeneracy occurs only rarely
in the Emery model. This has been checked by calculating exactly
the eigenvalues for CFs up to $n = 3$.

\begin{table}
\begin{tabular}{||l|l|l|l||}
Atom&   Ion&    Z& Radius       \\ \hline
Ytterbium&      Yb$^{3+}$&      70&     0.86      \\ \hline
Yttrium&        Y$^{3+}$&       39&     0.89      \\ \hline
Neodymium&      Nd$^{3+}$&      60&     0.995    \\ \hline
Lanthanum&      La$^{3+}$&      57&     1.02     \\
\end{tabular}
\caption[b]{\protect Ionic radius in \AA  ~and atomic number Z of some
trivalent rare earth ions RE$^{3+}$.}
\label{IONICRADII}
\end{table}

\begin{figure}[htb]
\caption[b]{\protect
Shown are two possible domains of the Ortho-I structure with
orientations of chains along the horizontal (a) and vertical (b) axes;
one of four domains of the Ortho-II
structure (c); and a typical pattern of the tetragonal configuration with
orientations of finite chain fragments along both $x$ and $y$ axes (d).
Squares and filled (open) circles denote Cu atoms and O atoms (vacancies),
respectively.}
\label{figure1}
\end{figure}

\begin{figure}[htb]
\caption[b]{\protect
Sketch of the ASYNNNI model, with interactions coupling
oxygen atoms. Squares and filled (open) circles symbolize
copper and oxygen atoms(vacancies).}
\label{figure2}
\end{figure}

\begin{figure}[htb]
\caption[b]{\protect
Schematic structural phase diagram of YBCO. Below the dashed curve, i.e.
below $T_R$, equilibrium can hardly be reached experimentally.}
\label{figure3}
\end{figure}

\begin{figure}[htb]
\caption[b]{\protect
Examples of simple perfect structures consisting of
the shortest CFs at $x=0.5$, (a) and (b).
The herringbone structure (a) would be favored by pure
electrostatic (quadrupolar) interactions \cite{schreck}.
Squares and filled (open) circles denote copper and oxygen atoms
(vacancies).}
\label{figure4}
\end{figure}

\begin{figure}[htb]
\caption[b]{\protect Free energy per oxygen
atom, $f$, of Cu-O chains versus chain
lenght (or number of oygen atoms), $n$,
for various values of $U_{\rm end}$ at
$T=0.06 \approx 900$ K. For each curve, the chemical potential has been
chosen so that the
concentration of oxygen holes is $0.3$ for the longest chains, independent
of $U_{\rm end}$. 
The free energy is measured
in units of the hopping parameter $t\approx 15000$ K.}
\label{Uend}
\end{figure}

\begin{figure}[htb]
\caption[b]{\protect Free energy per oxygen
atom, $f$, of Cu-O chains versus
length, $n$, for equidistant values of the chemical
potential, ranging from 1.5 to 2.0, at
$U_{\rm end} = 1.0$ and  $T=900$ K.}
\label{mu}
\end{figure}

\begin{figure}[htb]
\caption[b]{\protect Free energy per oxygen atom, $f$, of Cu-O chains
vs length, $n$, for a few
temperatures, with $U_{\rm end} = 1.0$ and the hole concentration
being 0.3
for the longest chains. Open symbols denote
fits to  expression (\ref{fn}).}
\label{FreeFit}
\end{figure}

\begin{figure}[htb]
\caption[b]{\protect $V_2$ vs temperature, $T$, for a
set of equidistant values of 
$U_{\rm end}$. The hole 
oxygen hole concentration is taken to be 0.3 for the longest, $n=6$, chains.} 
\label{FreeFitV2}
\end{figure}

\begin{figure}[htb]
\caption[b]{\protect $\widetilde{V}_2$ vs temperature, $T$,
for a set of equidistant values of
$U_{\rm end}$. The oxygen hole concentration is 0.3 for the longest chains.} 
\label{FreeFitV2tilde}
\end{figure}

\begin{figure}[htb]
\caption[b]{\protect Ortho-II order parameter, $m_{\rm II}$,
vs $U_{\rm end}$, at $T_{\rm II}=0.03 $ corresponding to 450 K. The full
symbols denote Monte Carlo data for the extended ASYNNNI model, applied
to YBCO; the 
interpolating line is included as a guide to the eye.}
\label{3DOII}
\end{figure}

\begin{figure}[htb]
\caption[b]{\protect $V_2$ (full symbols) and $\widetilde{V}_2$ (open
symbols) vs chemical potential $\mu$ for various temperatures, at 
$U_{\rm end} = 1.02$.
}
\label{MUVARIV2}
\end{figure}

\begin{figure}[htb]
\caption[b]{\protect Illustration of shifting of bands by going from
the semiconducting (tetragonal) to the metallic (orthorhombic)
state in REBCO compounds.
The 1D hole band filling is larger in the semiconducting state.
}
\label{figure_12}
\end{figure}

\begin{figure}[htb]
\caption[b]{\protect Phase diagram (temperature $T(K)$ in Kelvin
vs. oxygen content $x$) obtained from MC simulations of the
extended ASYNNNI model for YBCO. Typically, systems with
40 $\times$ 40 copper ions were considered.}
\label{PHASEDIAGRAM}
\end{figure}

\begin{figure}[htb]
\caption[b]{\protect Concentration of twofold coordinated Cu ions, $c_2$,
versus oxygen content, $x$, at various temperatures.
Simulations of the extended ASYNNNI model for YBCO were done at $T=300$ K and 
600 K. MC data are shown by full dots; interpolating 
lines are included, for clarity,
as guides to the eye. Experimental results
\cite{tranq,tolen} are denoted by open symbols.}
\label{CU2}
\end{figure}

\begin{figure}[htb]
\caption[b]{\protect Concentration of two-, three- and fourfold coordinated
Cu ions, $c_l$ ($l$ =2,3, and 4) versus oxygen content, $x$, at $T = 450$ K,
simulating the extended ASYNNNI model for YBCO,
$\mu$ = 1.59 (full symbols and, as guide
to the eye, lines). For comparison, experimental data
\cite{luetg1} are shown (open symbols).}
\label{CUfoldY}
\end{figure}

\begin{figure}[htb]
\caption[b]{\protect Notation as in Figure 15; however, in this
case for NdBCO, 
taking the chemical potential $\mu$ to be 1.62.}
\label{CuFoldNd}
\end{figure}

\begin{figure}[htb]
\caption[b]{\protect Notation as in Figure 15; however, in this
case for LaBCO,
taking the chemical potential $\mu$ to be 1.66.}
\label{CuFoldLa}
\end{figure}

\begin{figure}[htb]
\caption[b]{\protect Number of holes per oxygen atom, $\overline{\nu_h(n)}/n$,
leaving a CF of length $n$ ($n=2,\dots,6$), as function
of temperature $T$, at
$\mu=1.59$ ($\nu_h=0$ for $n=1$).}
\label{ch_tr}
\end{figure}

\begin{figure}[htb]
\caption[b]{\protect Charge transfer, $n_h$, as
function of oxygen content $x$, at
various temperatures from 300 K to 600 K, simulating
the extended ASYNNNI model 
for YBCO.}
\label{CHARGETRANSFER}
\end{figure}

\begin{figure}[htb]
\caption[b]{\protect Charge transfer, $n_h$, as
function of oxygen content, $x$,
simulating the extended ASYNNNI model mimicing RE= La, Nd, Y and Yb 
(from bottom to top).}
\label{CHARGETRANSFERall}
\end{figure}

\begin{figure}[htb]
\caption[b]{\protect Average chain length, $\ell$, as
function of oxygen content, $x$,
simulating the extended ASYNNNI model mimicing RE= La, Nd, and Y
(from bottom to top).
Full and open symbols denote direct and indirect (through $c_3$)
determination of $\ell$, respectively; see text.}
\label{CHAINLENGTH}
\end{figure}


\begin{references}
\bibitem[*]{gu} On leave from Landau Institute for Theoretical Physics,
Chernogolovka, Moscow District 142432, Russia.
\bibitem{luetg1} H. L\"utgemeier, I. Heinmaa, D. Wagener, and S.M. Hosseini,
in "Phase Separation in Cuprate Superconductors", Proc. of the
International Workshop in Cottbus 1993, p. 225. Edited by E. Sigmund and
K.A. M\"uller, (Springer Verlag 1994).
\bibitem{luetg2} H. L\"utgemeier, S. Schmenn, P. Meuffels, O. Storz,
R. Sch\"ollhorn, Ch. Niedermayer, I. Heinmaa, and Yu. Baikov, Physica C
{\bf 267}, 191 (1996).
\bibitem{tranq} J.M. Tranquada, S.M. Heald, A.R. Moodenbaugh, and Y. Xu,
Phys.Rev. B {\bf 38}, 8893 (1988).
\bibitem{tolen} H. Tolentino, F. Baudelet, A. Fontaine,  T. Gourieux,
G. Krill, J.Y. Henry, and J. Rossat-Mignod, Physica C {\bf 192}, 115 (1992).
\bibitem{krol} A. Krol, Z.H. Ming, Y.H. Kao, N. N\"ucker, G. Roth, J. Fink,
G.C. Smith, K.T. Park, J. Yu, A.J. Freeman, A. Erband, G. M\"uller-Vogt,
J. Karpinski, E. Kaldis, and K. Sch\"onmann, Phys.Rev. B {\bf 45},
2581 (1992).
\bibitem{BHHKW90} M. Buchgeister, W. Hiller, S.M. Hosseini, K. Kopitzki, and
D. Wagener in "Physics and Material Science of High Temperature
Superconductors", p. 319. Edited by R. Kossowsky, S. Methfessel and 
D. Wohlleben (Kluwer Academic Publishers, The Netherlands, 1990).
\bibitem{veal_4} B.W. Veal, A.P. Paulikas, J.W. Downey, H. Claus, 
K. Vandervoort, G. Tomblins, H. Shi, M. Jensen, and L. Morss, Physica C 
{\bf 162}--{\bf 164}, 97 (1989).
\bibitem{lindemer} T.B. Lindemer, B.C. Chakoumakos, E.D. Specht, R.K. Williams,
and Y.J. Chen,  Physica C {\bf 231}, 80 (1994).
\bibitem{lant_1} Y. Tobura, in ``Physics of High-Temperature Superconductors'',
Springer Series in Solid-State Sciences, V. 106, p. 191. Edited by S. Maekawa
and M. Sato (Springer-Verlag, 1992).
\bibitem{lant_2} The high-T$_{\rm c}$ superconductors
La$_{2-x}$Sr$_x$CuO$_4$ for which $T_c(x)$
exhibits a maximum in $T_c(x)$ at $x\approx 0.15 - 0.2$.
\bibitem{veal_1} H. Claus, S. Yang, A.P. Paulikas, J.W. Downey, and B.W. Veal,
Physica C {\bf 171}, 205 (1990).
\bibitem{gant} G.V. Uimin, V.F. Gantmakher, A.M. Neminsky, L.A. Novomlinsky,
D.V. Shovkun, and P. Br\"ull, Physica C {\bf 192}, 481 (1992).
\bibitem{zeis} Th. Zeiske, R. Sonntag, D. Hohlwein, N.H. Andersen, and Th. Wolf,
Nature {\bf 353}, 542 (1991).
\bibitem{plakh_1} P. Burlet, V.P. Plakhty, C. Martin, and J.Y. Henry,
Phys.Lett. A {\bf 167}, 401 (1992).
\bibitem{plakh_2} V. Plakhty, A. Stratilatov, Yu. Chernenkov, V. Fedorov,
S.K. Sinha, C.K. Loong, B. Gaulin, M. Vlasov, and S. Moshkin, 
Solid State Commun. {\bf 84}, 639 (1992).
\bibitem{hohl_2} D. Hohlwein, in Proc. of the International School of
Crystallography in Erice 1993, edited by E. Kaldis, (Kluwer, Dordrecht 1994).
\bibitem{and_1} H.F. Poulsen, N.H. Andersen, J.V. Andersen, H. Bohr, and
O.G. Mouritsen, Nature {\bf 343}, 544 (1990).
\bibitem{gaw_2} P. Gawiec, D. R. Grempel,
G. Uimin, and J. Zittartz, Phys.Rev. B {\bf 53}, 5880 (1996).
\bibitem{font_1} D. de Fontaine, L.T. Wille, and S.C. Moss, Phys.Rev. B
{\bf 36}, 5709 (1987).
\bibitem{wille} L.T. Wille and D. de Fontaine, Phys.Rev. B {\bf 37}, 2227
(1988).
\bibitem{font_2} L.T. Wille, A. Berera, and D. de Fontaine, Phys.Rev.Lett.
{\bf 60}, 1065 (1988); R. Kikuchi and J.S. Choi, Physica C {\bf 160}, 347
(1989).
\bibitem{OLDSET2} G. Ceder, M. Asta, W. C. Carter, M. Kraitchman,
D. de Fontaine, M. E. Mann and M. Sluiter, Phys.Rev. B {\bf41},  8698 (1990).
\bibitem{selk} W. Selke and G. Uimin, Physica C {\bf 214}, 37 (1993).
\bibitem{fiig1} T. Fiig, J.V. Andersen, N.H. Andersen, P.A. Lindg\aa rd, O.G.
Mouritsen, and H.F. Poulsen, Physica C {\bf 217}, 34 (1993).
\bibitem{fiig2} T. Fiig, N.H. Andersen, J. Berlin, and P.A. Lindg\aa rd,
Phys.Rev. B {\bf 51}, 12246 (1995).
\bibitem{kinet} J.V. Andersen, H. Bohr, and O.G. Mouritsen, Phys.Rev. B
{\bf 42}, 283 (1990); H.F. Poulsen, N.H. Andersen, J.V. Andersen, H. Bohr, and
O.G. Mouritsen, Phys.Rev.Lett. {\bf 66}, 465 (1991).
\bibitem{eins} N.C. Bartelt, T.L. Einstein, and L.T. Wille, Phys.Rev. B {\bf
40},
10759 (1989); T. Aukrust, M.A. Novotny, P.A. Rikvold, and D.P. Landau,
Phys.Rev. B {\bf 41}, 8772 (1990); D.K. Hilton, B.M. Gorman, P.A. Rikvold, and
M.A. Novotny, Phys.Rev. B {\bf 46}, 381 (1992); D.J. Liu, T.L. Einstein,
P.A. Sterne, and L.T. Wille, Phys.Rev. B {\bf 52}, 9784 (1995).
\bibitem{oit} J. Oitmaa, Y. Jie and  L.T. Wille, J.Phys.: Condens. Matter
{\bf 5}, 4161 (1993).
\bibitem{uim_int} G. Uimin, Int.J.Mod.Phys. B {\bf 6}, 2291 (1992).
\bibitem{uim_pr} G. Uimin, Phys.Rev. B {\bf 50}, 9531 (1994).
\bibitem{veal_3} S. Yang, H. Claus, B.W. Veal, R. Wheeler, A.P. Paulikas, and
J.W. Downey, Physica C {\bf 193}, 243 (1992).
\bibitem{OIITc} P. Gerdanian and C. Picard, Physica C {\bf 204},  419 (1993).
\bibitem{schw} W. Schwarz, O. Blaschko, G. Collin, and F. Marucco,
Phys.Rev. {\bf B 48}, 6513 (1993).
\bibitem{alig_ssc} A.A. Aligia, Solid State Commun. {\bf 78}, 739 (1991).
\bibitem{alig} A.A.Aligia, Europhys. Lett. {\bf 26}, 153 (1994).
\bibitem{yakh} F. Yakhou, V. Plakhty, G. Uimin, P. Burlet,
B. Kviatkovsky, J.Y. Henry, J.P. Lauriat, E. Elkaim, and E. Ressouche,
Solid State Commun. {\bf 94}, 695 (1995).
\bibitem{schl} P. Schleger, W.N. Hardy, and B.X. Yang, Physica C {\bf 176}
261 (1991); P. Schleger, W.N. Hardy, and H. Casalta, Phys.Rev. B {\bf 49}, 514
(1994).
\bibitem{alar} M.A. Alario-Franco, C. Chaillout, J.J. Capponi, J. Chenavas, and
M. Marezio, Physica C {\bf 156}, 455 (1988).
\bibitem{rey} J. Reyes-Gasga, T. Krelels, G. van Tendeloo, J. van Landuyt,
S. Amelinckx, X.H.M. Bruggink, and H. Verweij, Physica C {\bf 159}, 831 (1989).
\bibitem{hohl} Th. Zeiske, D. Hohlwein, R. Sonntag, F. Kubanek, and G. Collin,
Z.Phys. B {\bf 86}, 11 (1992).
\bibitem{schreck} M.Schreckenberg and G.Uimin, Sov.Phys. JETP Letters {\bf 58},
143 (1993).
\bibitem{rm} G. Uimin and J. Rossat-Mignod, Physica C {\bf 199}, 251 (1992).
\bibitem{emer} V.J. Emery, Phys.Rev.Lett. {\bf 58}, 2794 (1987).
\bibitem{garc} A.A.Aligia and J. Garc\'es, Phys. Rev. {\bf 49}, 524 (1994).
\bibitem{gaw_1} P. Gawiec, D.R. Grempel, A.-C. Riiser, H. Haugerud,
and G. Uimin, Phys.Rev. B {\bf 53}, 5872 (1996).
\bibitem{haug} H. Haugerud, F. Ravndal, and G. Uimin, J.Phys.: Condens. Matter
{\bf 5}, 6895 (1993).
\bibitem{lan} C. Lanczos, J. Res. Nat. Bur. Standards {\bf 45}, 255 (1950).
\bibitem{STERNE89} P.A. Sterne and L.T. Wille, Physica C {\bf 162}, 223
(1989).
\bibitem{comment}
The metallic state is usually associated with orthorhombicity with an 
increased lattice constant in the chain direction.
This reduces the hopping amplitude $t$ and narrows the 1D band.
The chemical potential counted from the 1D bottom determines the
1D band filling, decreasing with $t$, see Figure \ref{figure_12}. 
The effective changes can be described by renormalizing
the chemical potential and temperature.
\bibitem{TORR89} J.B. Torrance, A. Benzinge, A.I. Nazzal, T.C. Huang,
S.P. Parkin, D.T. Keane, S.J. LaPlaca, P.M. Horn, and G.A. Held,
Phys.Rev. B {\bf 40}, 8872 (1989).
\bibitem{SHAFER89} M.W. Shafer, T. Penney, B.L. Olson, R.L. Greene,
and R.H. Koch, Phys. Rev. B {\bf 39}, 2914 (1989).
\bibitem{BHB92} G. Baumg\"artel, W. H\"ubner and K.H. Bennemann, Phys. Rev. B
{\bf 45}, 308 (1992).
\bibitem{Rotter91} L.D. Rotter, Z. Schlesinger, R.T. Collins, F. Holtzberg,
C. Field, U.W. Welp,
G.W. Crabtree, J.Z. Liu, Y. Fang, K.G. Vandervoort, and S. Fleshler, Phys. Rev.
Lett. {\bf 67}, 2741 (1991).
\bibitem{Whangbo91} M.-H. Whangbo and C.C. Toradi, Science {\bf 249}, 1143
(1990).
\bibitem{VEAL91} B.W. Veal and A.P. Paulikas, Physica C {\bf 184}, 321 (1991).
\end{references}
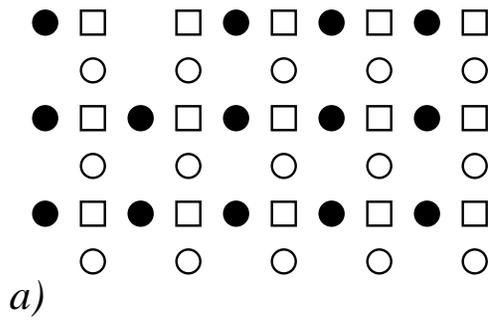
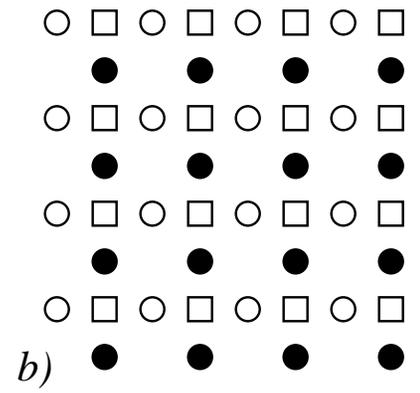
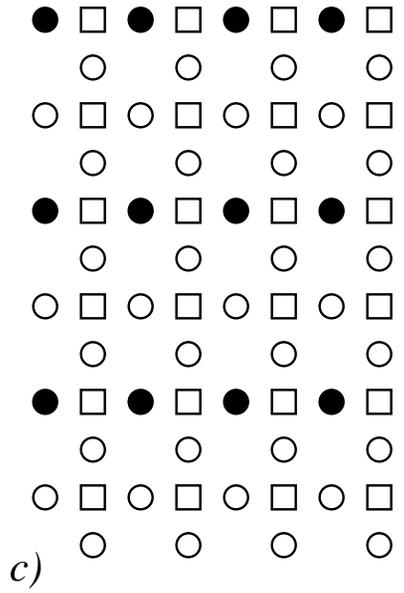
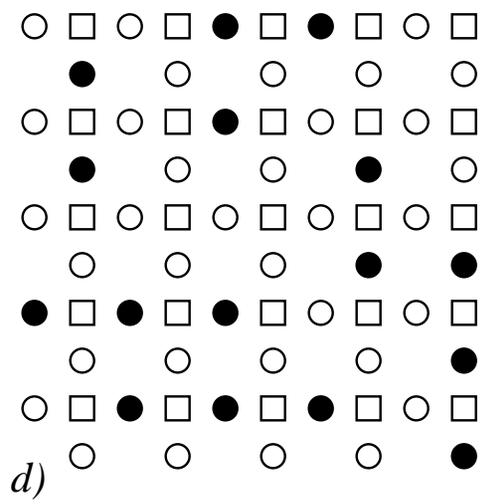
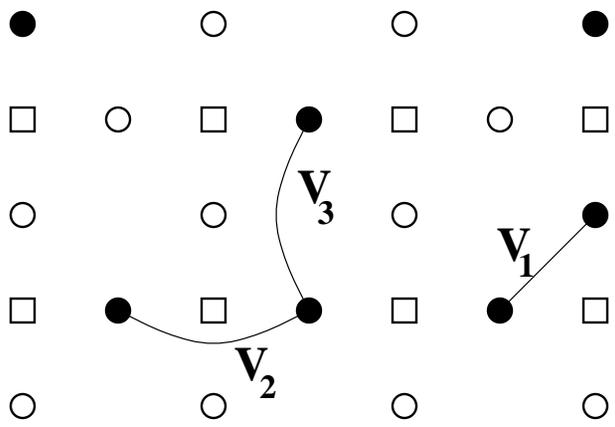
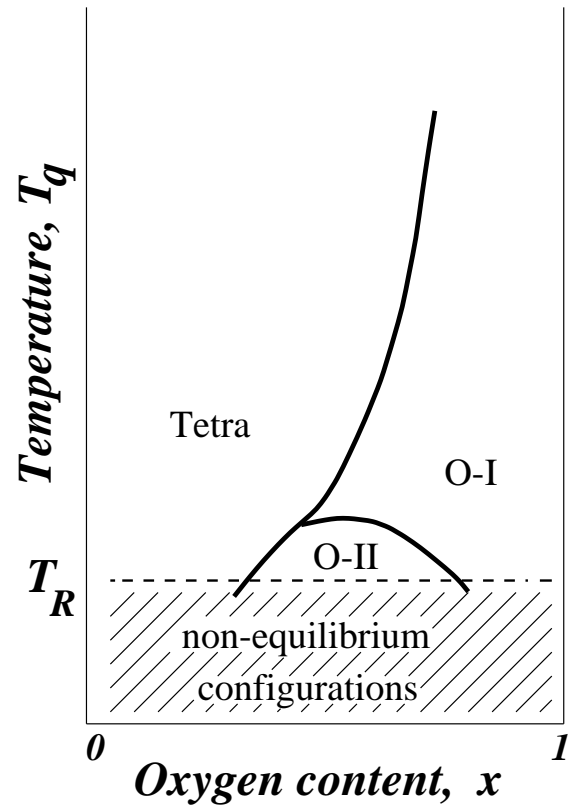
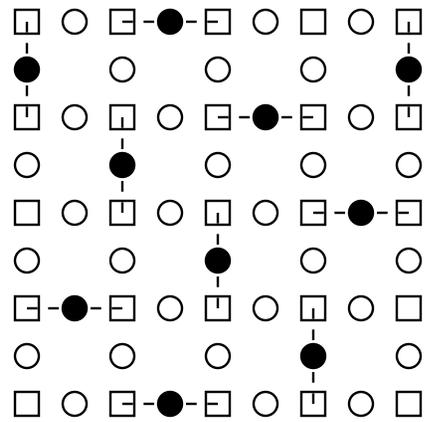
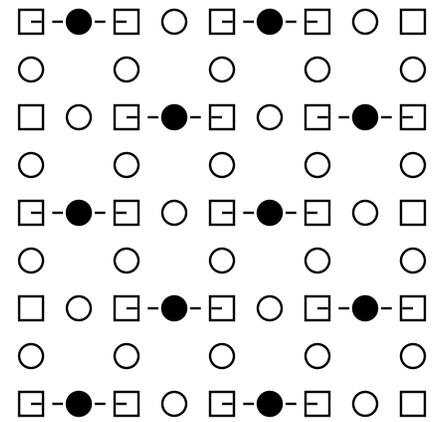
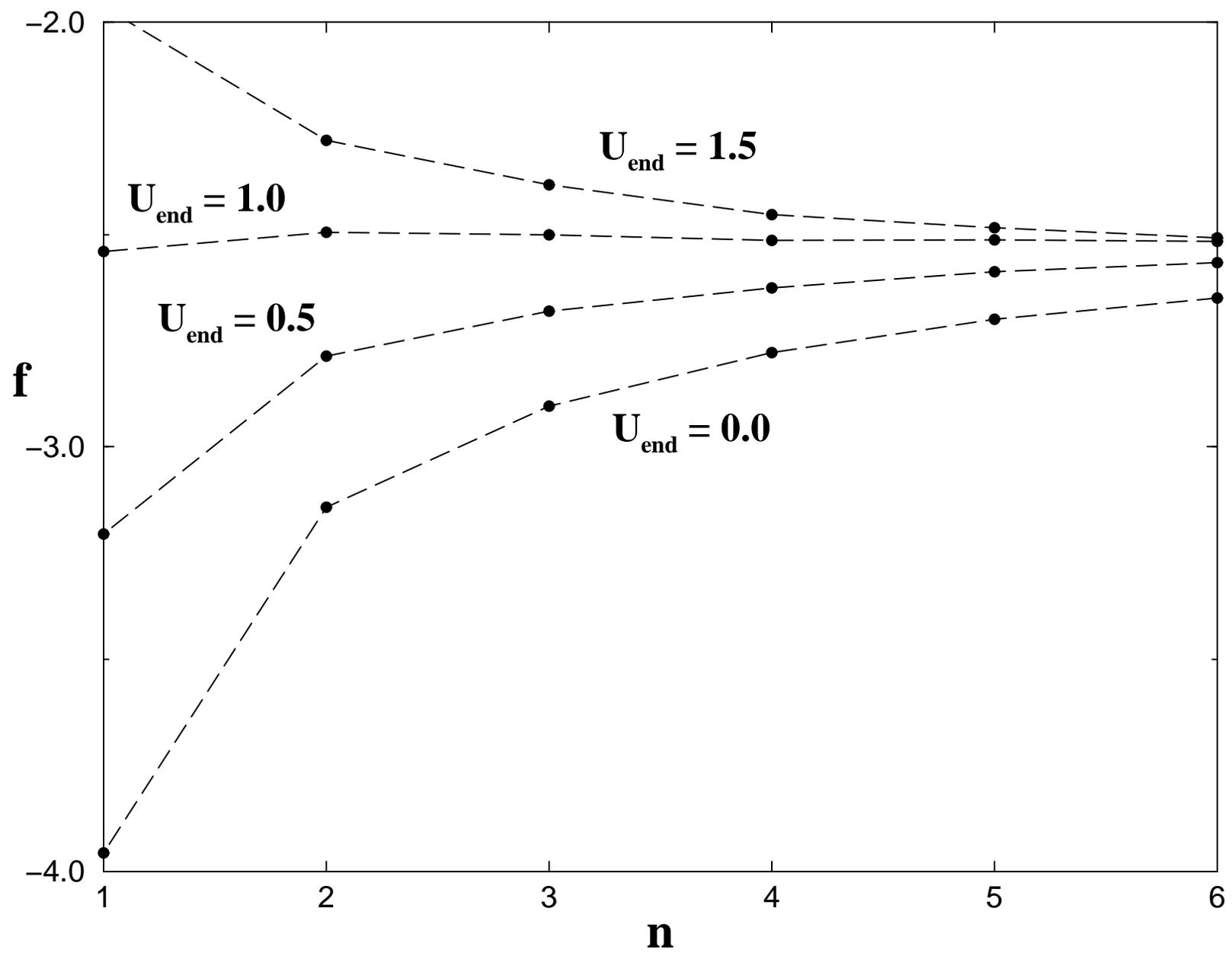
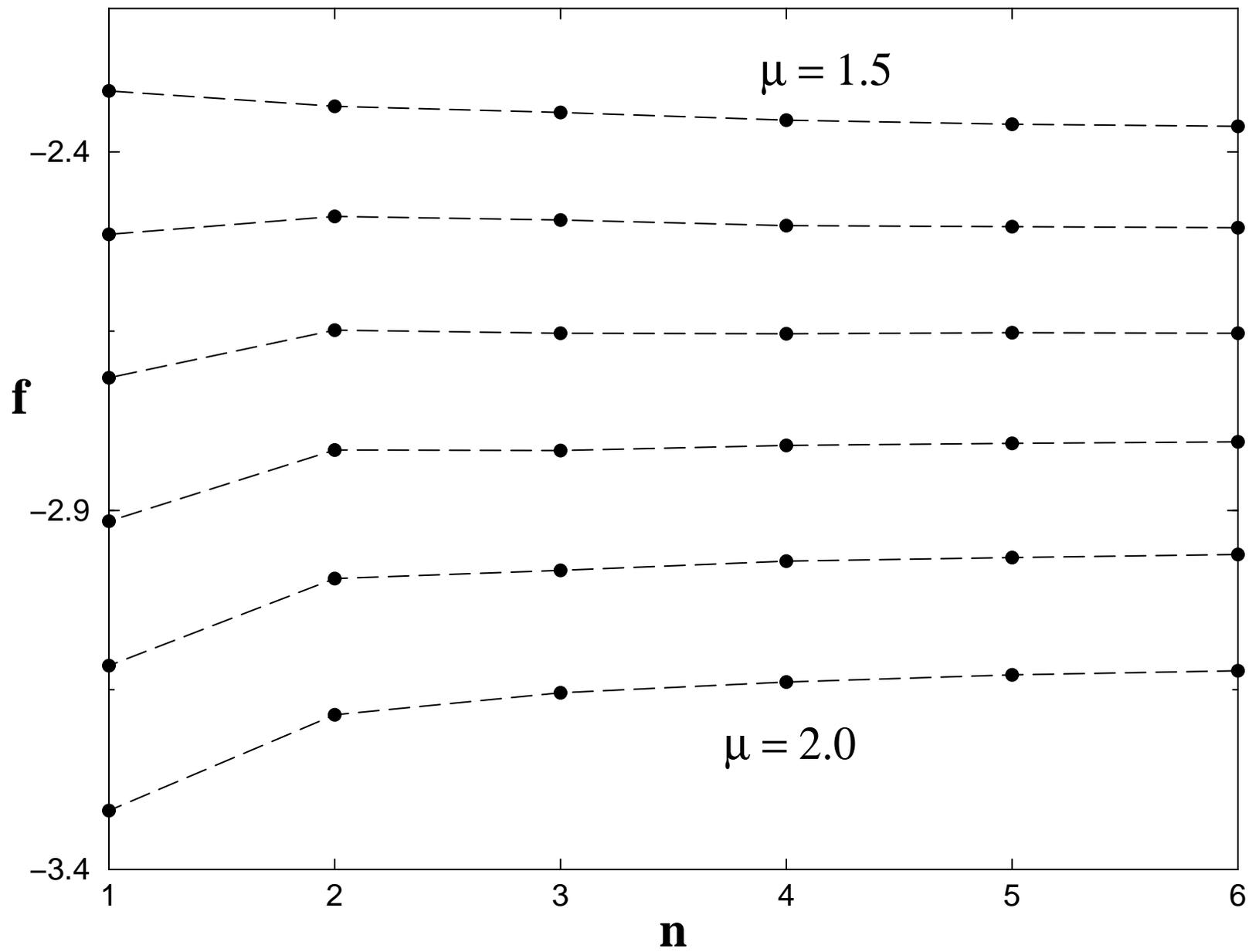
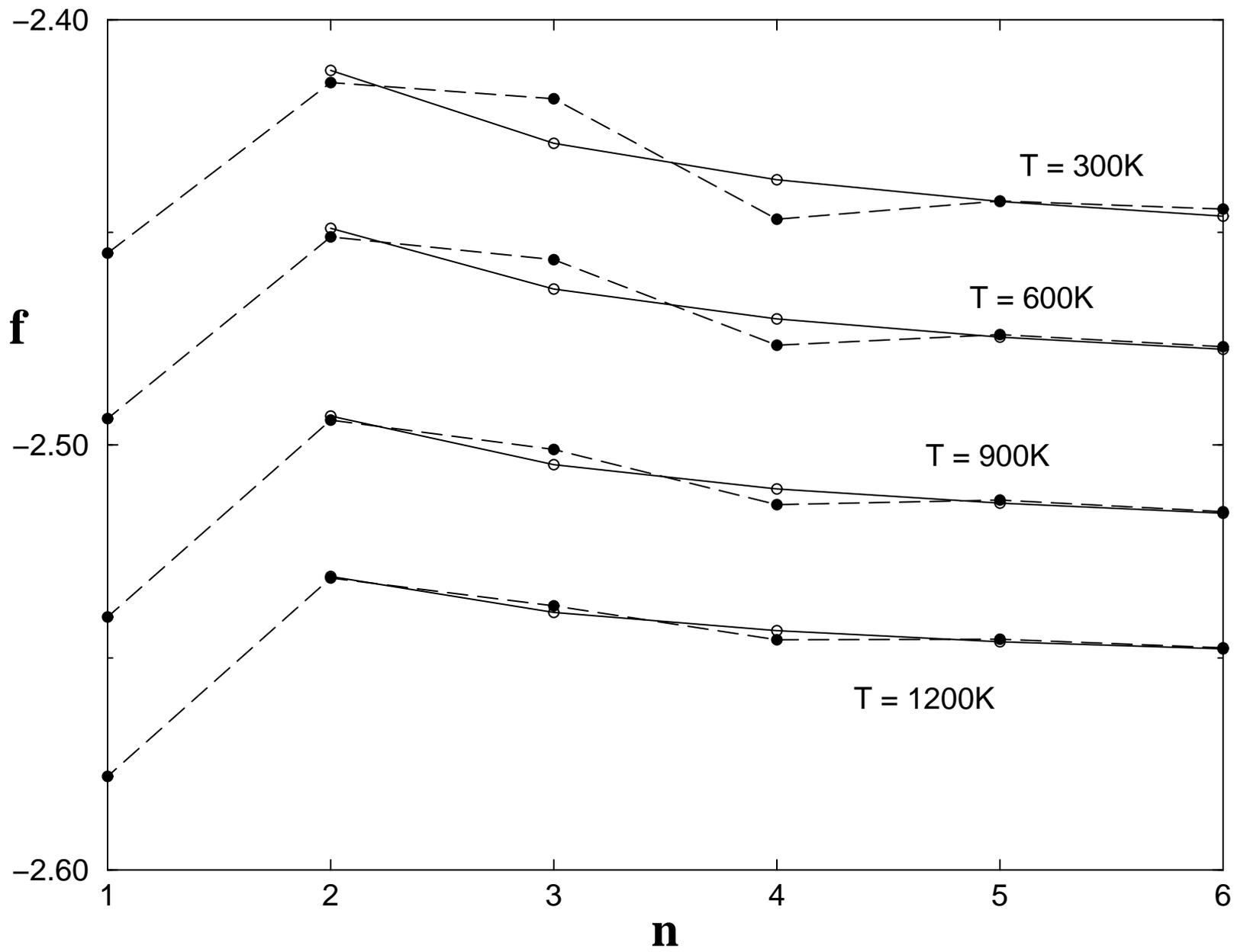
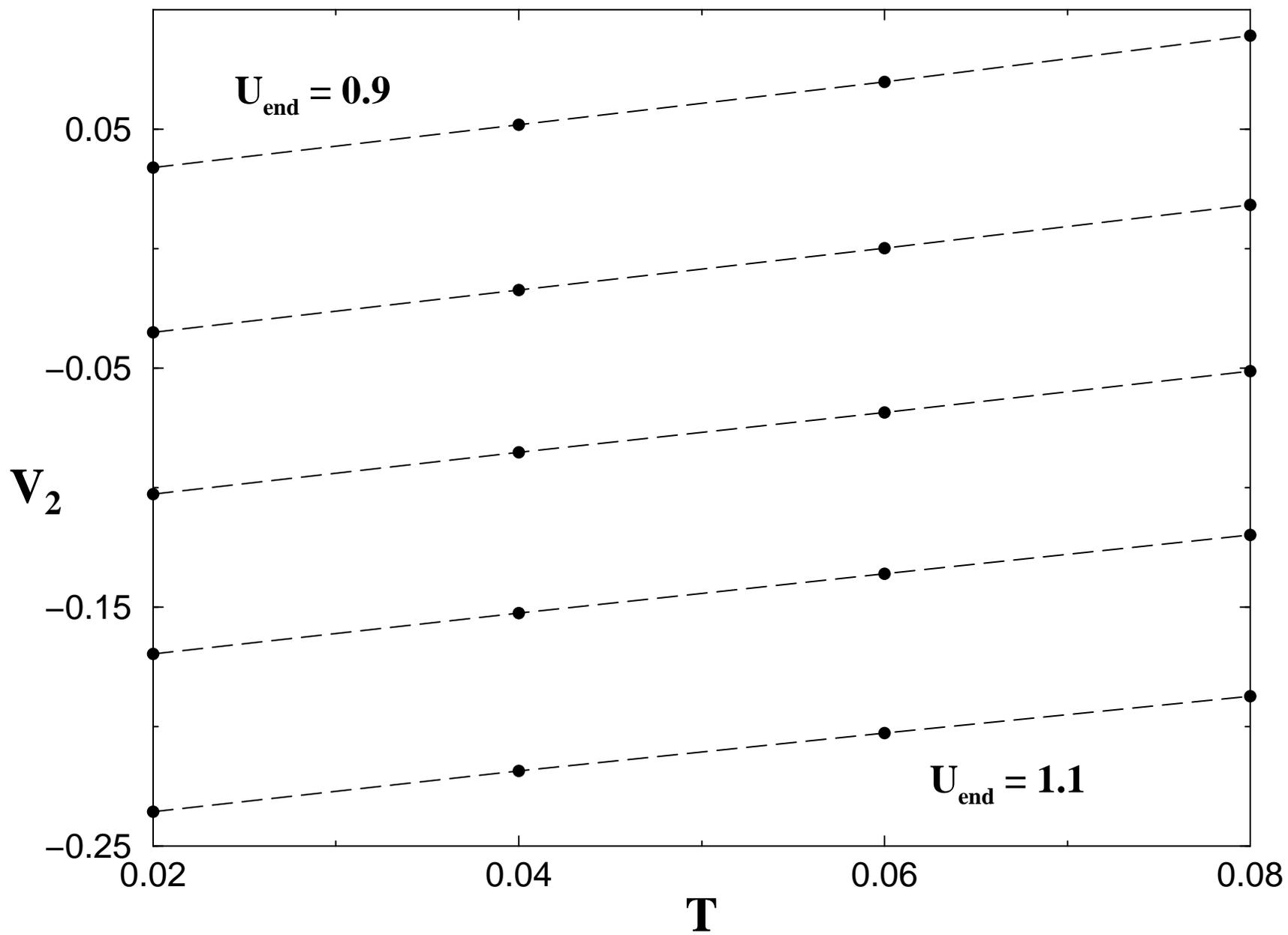
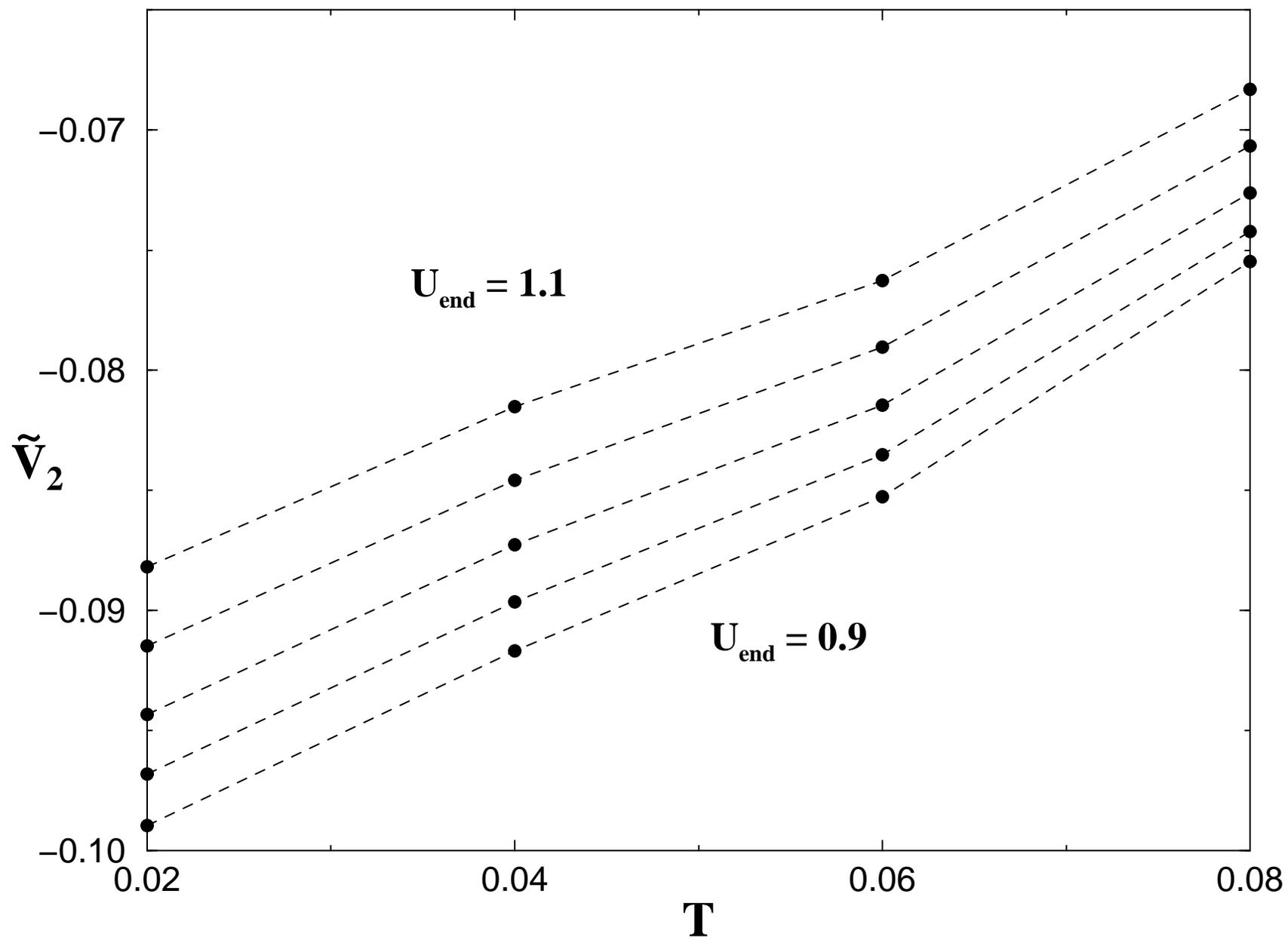
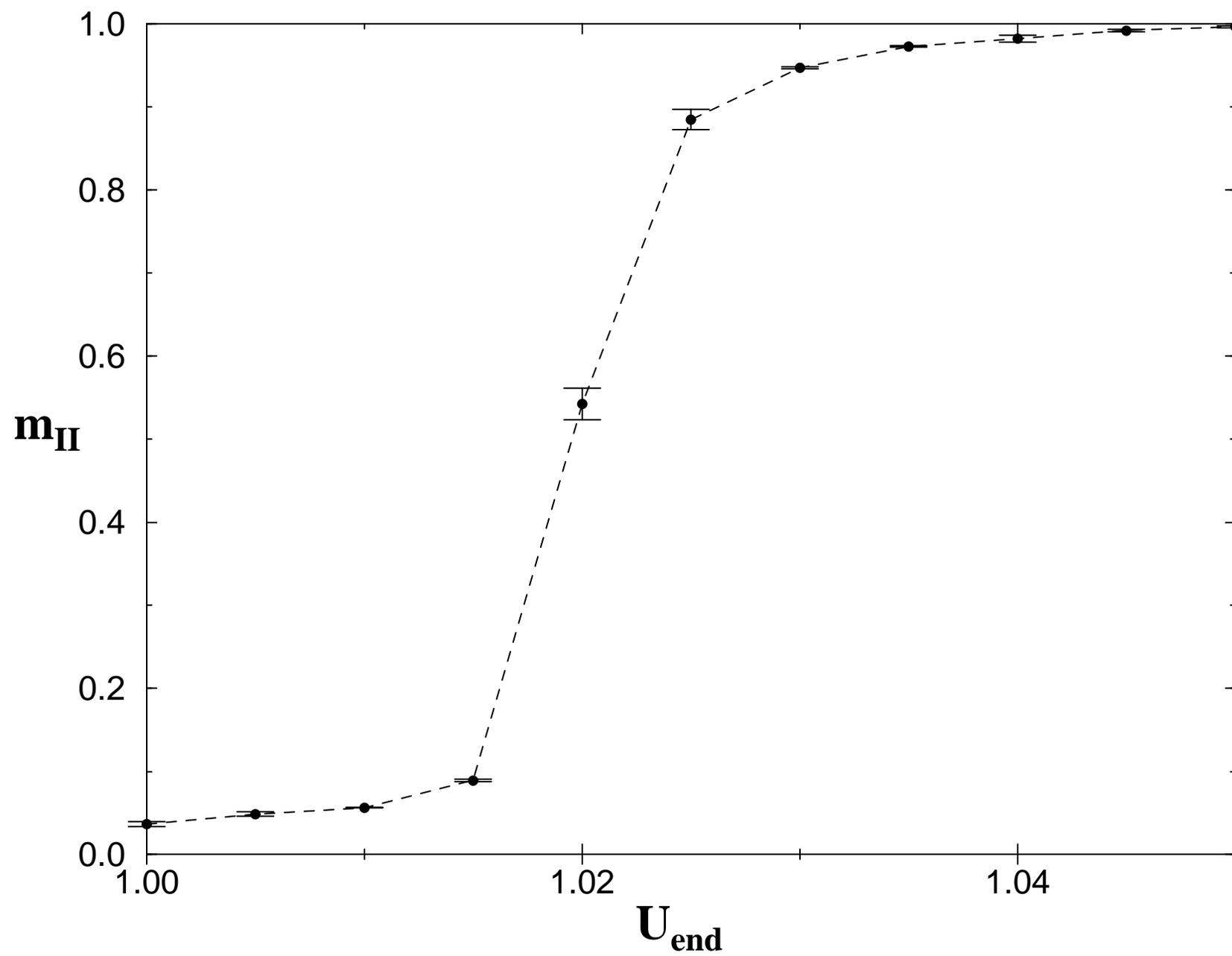
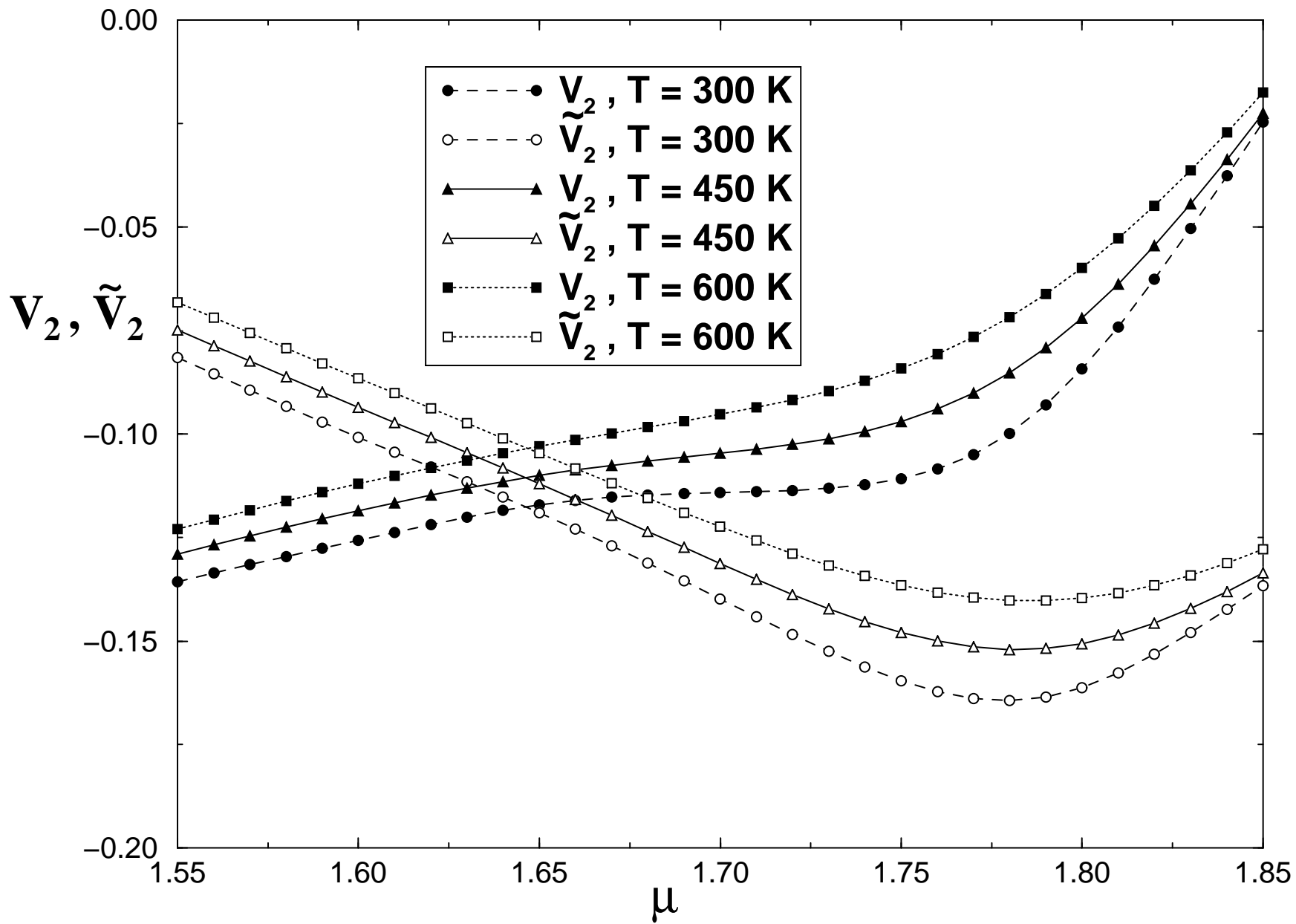
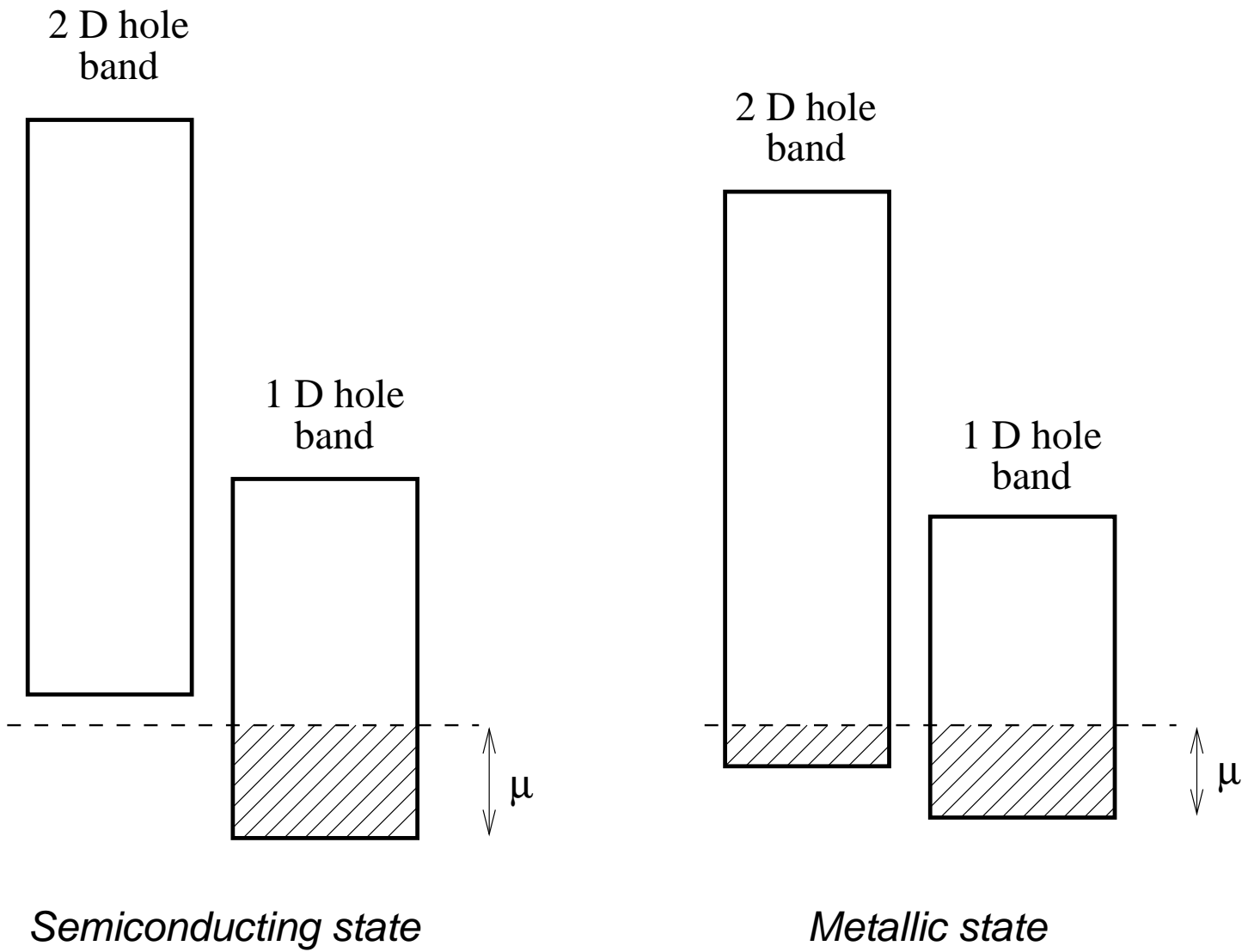
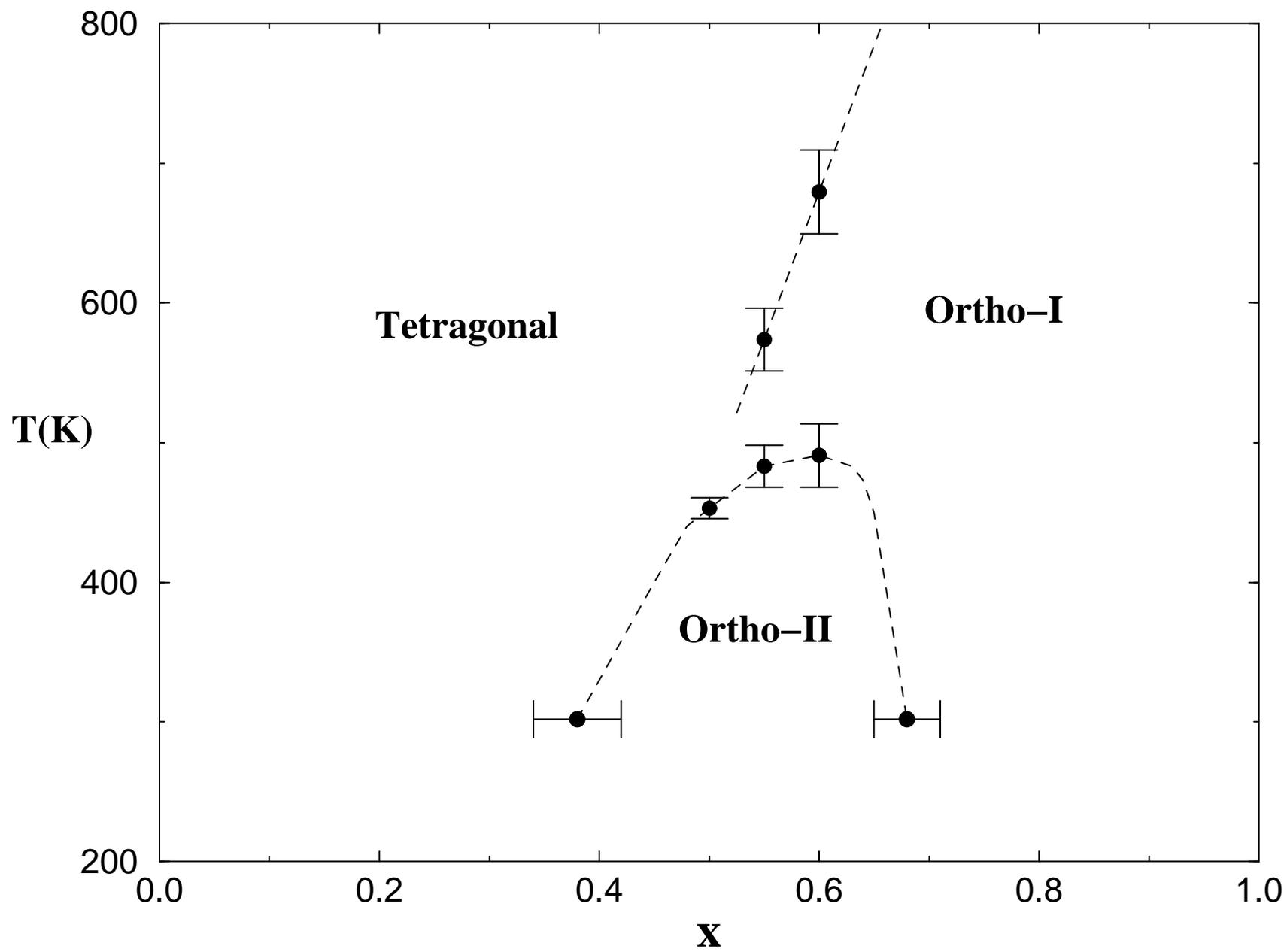
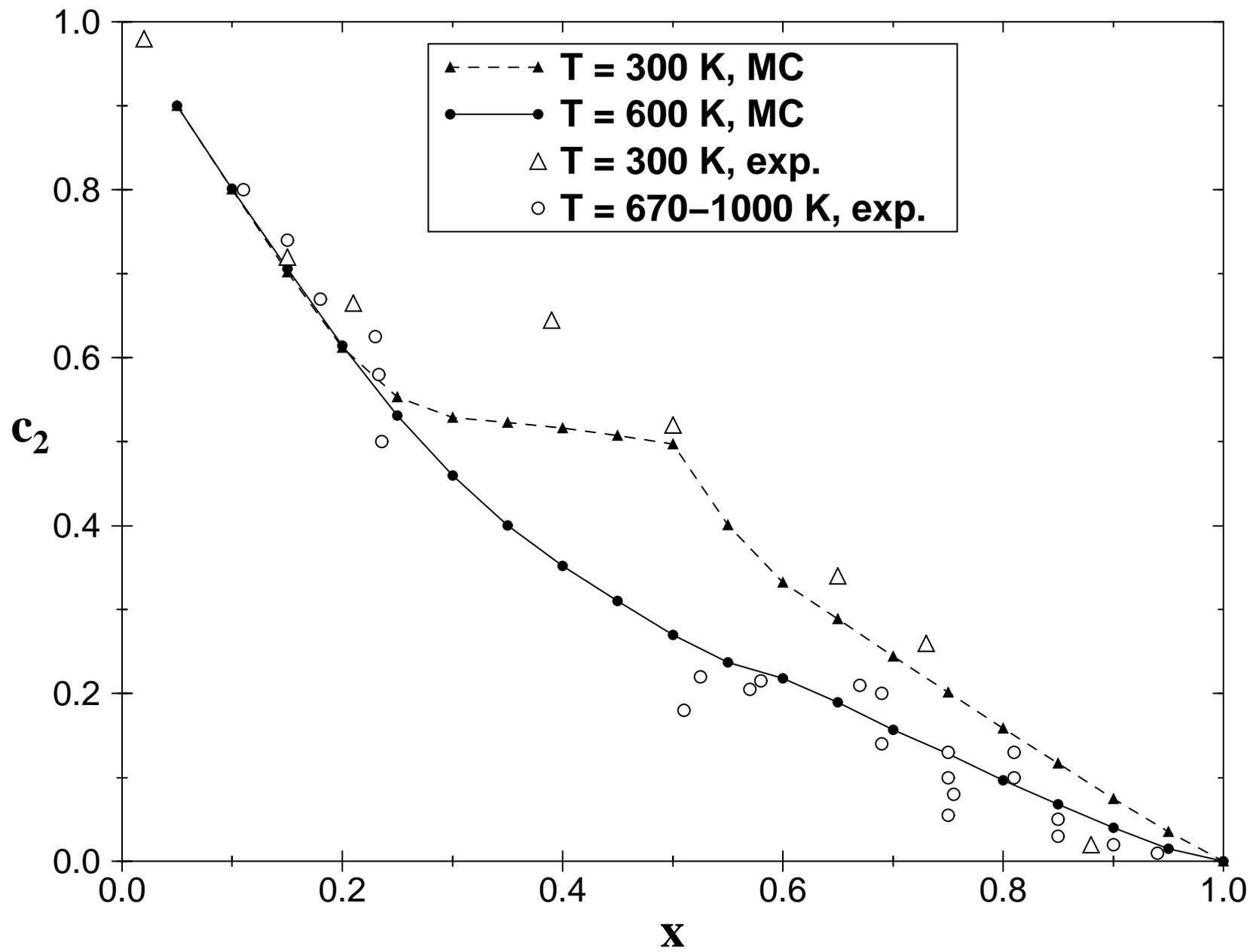
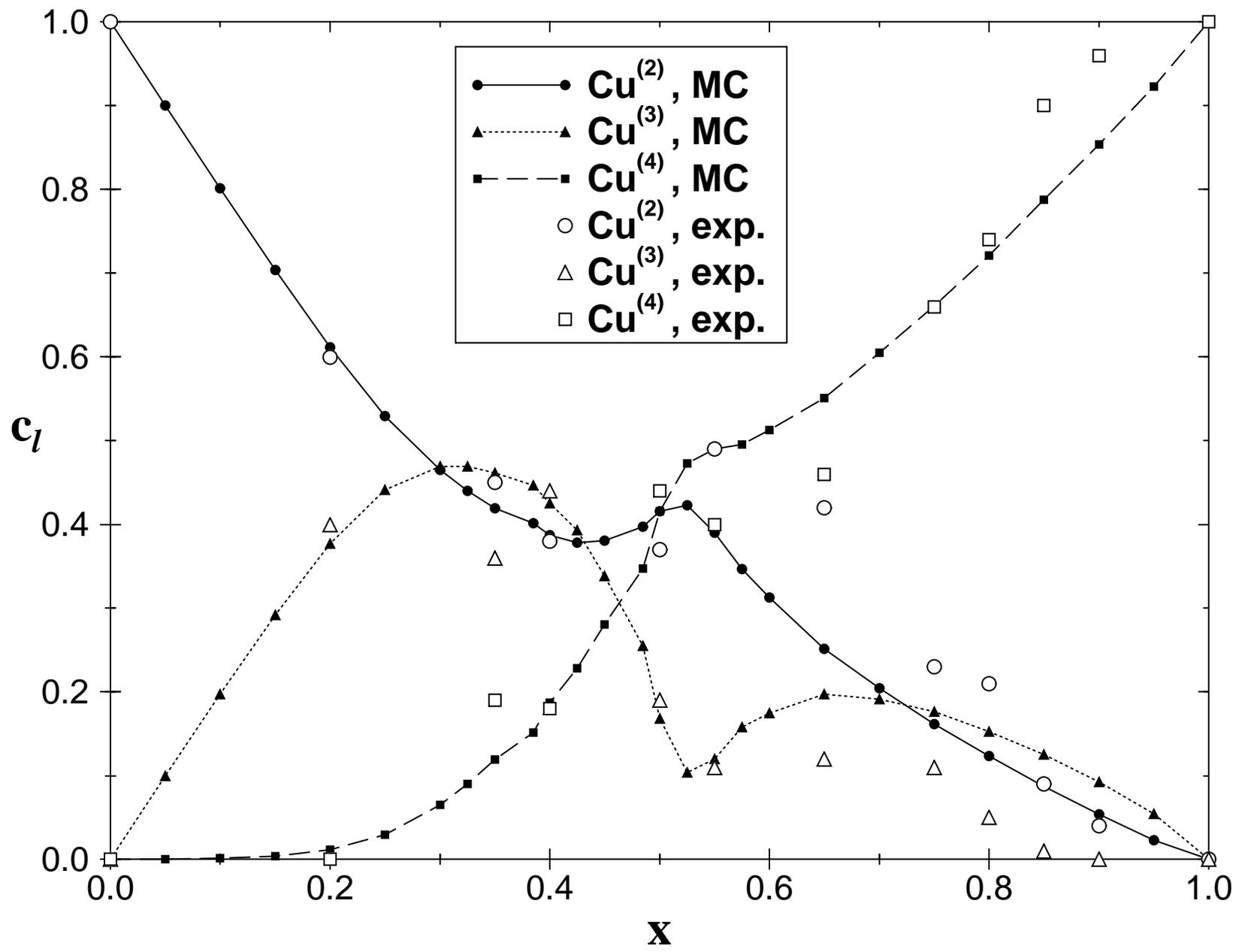
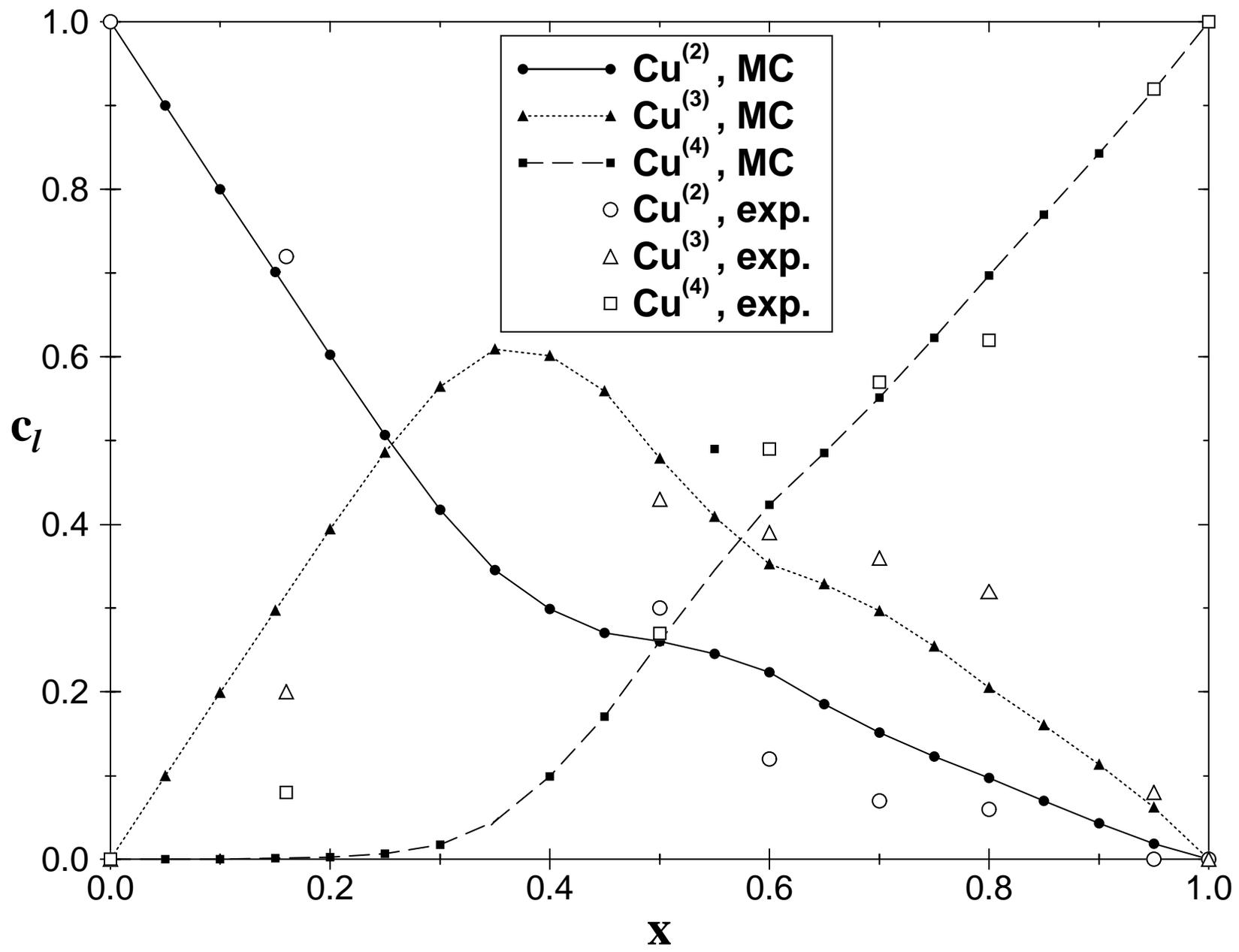
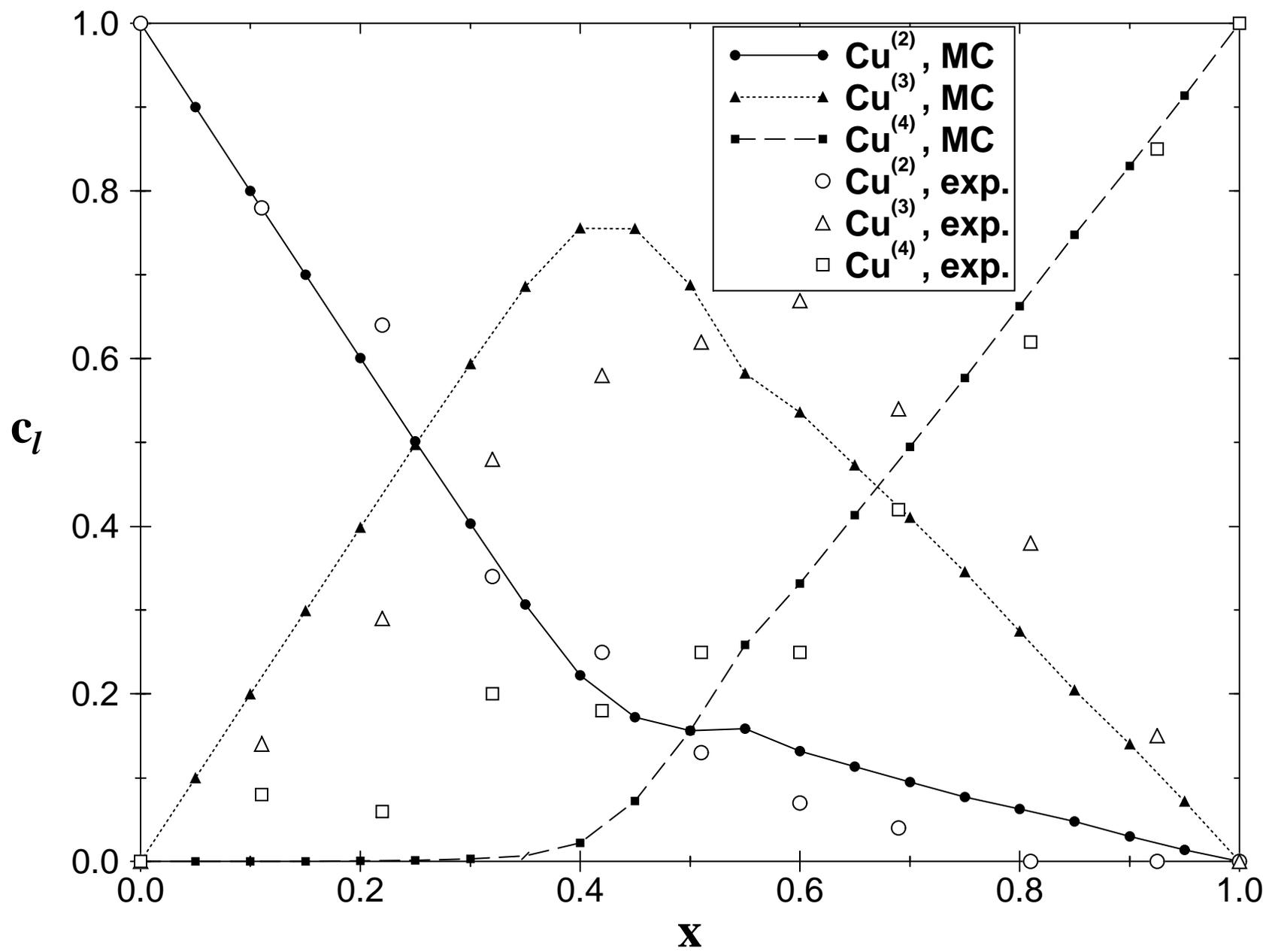
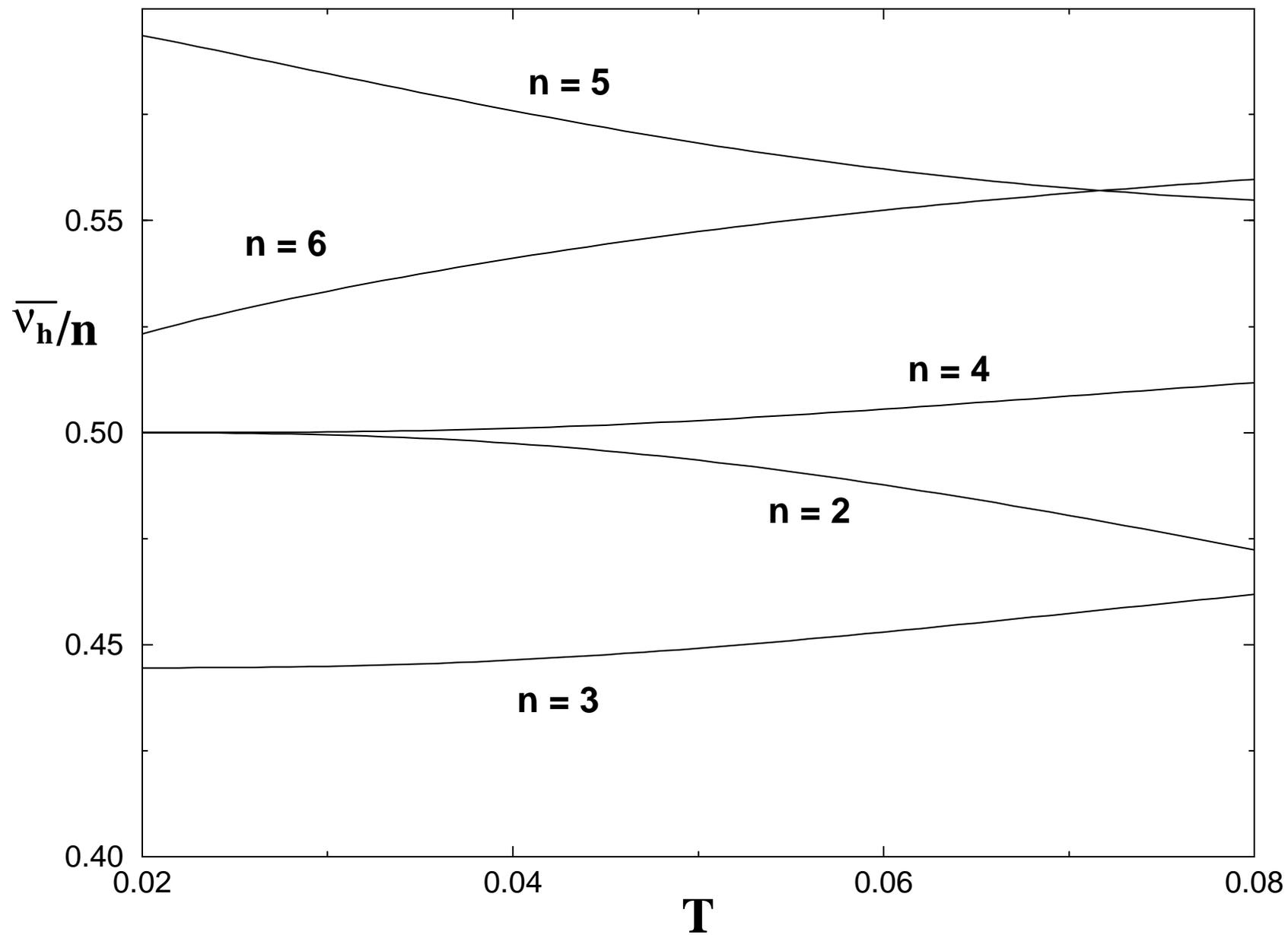
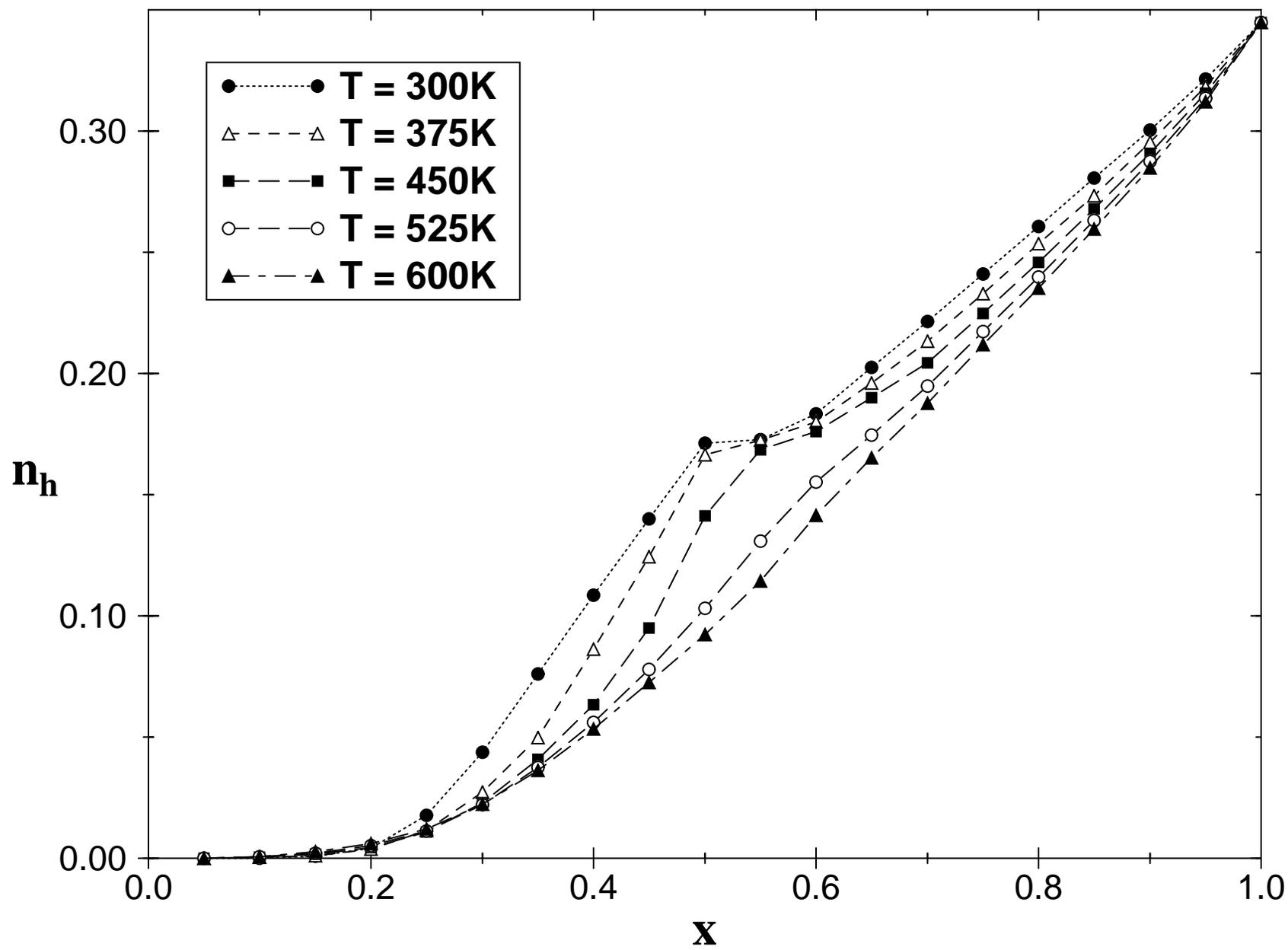
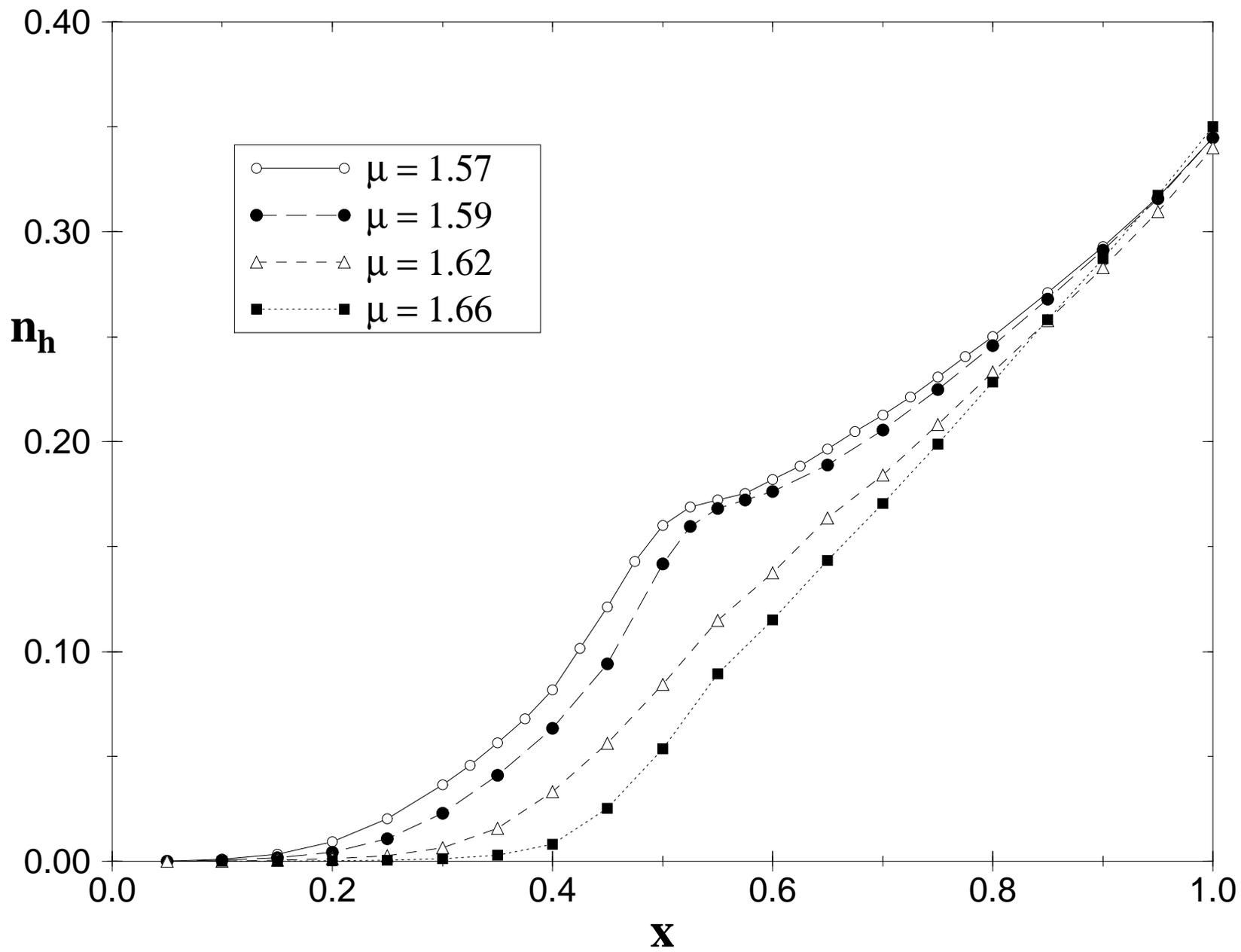
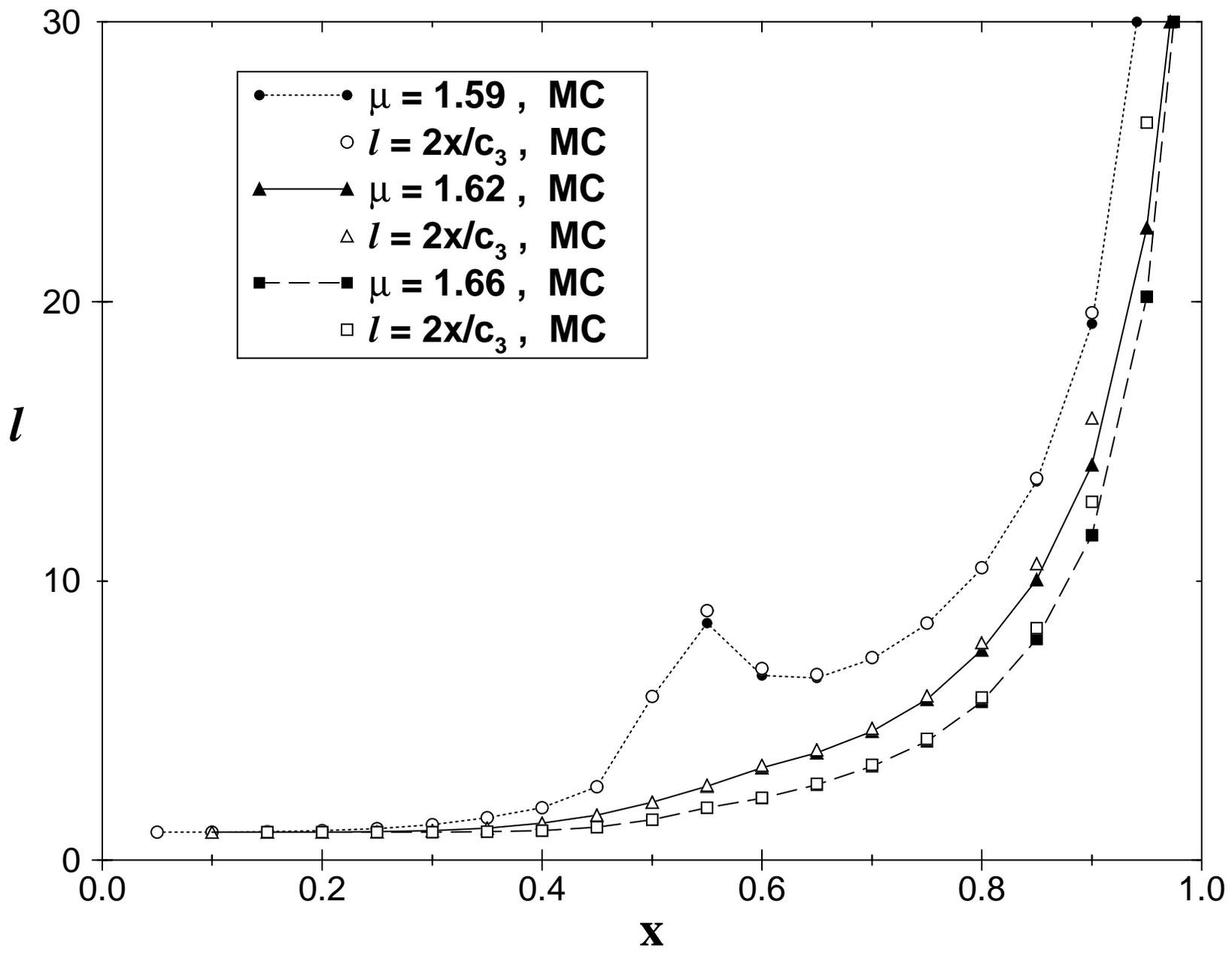
\end{document}